\documentclass[aps,showpacs,floats,prb,amsmath,amssymb,twocolumn]{revtex4}
\usepackage{graphicx}
\usepackage{color}

\begin{document}

\title{Thermal drag revisited: Boltzmann versus Kubo}

\author{Suhas Gangadharaiah}
\affiliation{Department of Physics, University of California, Irvine,
California 92697, USA}
\author{A. L. Chernyshev}
\affiliation{Department of Physics, University of California, Irvine,
California 92697, USA}
\author{Wolfram Brenig}
\affiliation{Institut f\"ur Theoretische Physik,
Technische Universit\"at Braunschweig, 38106 Braunschweig, Germany}
\date{\today}
\begin{abstract}
The effect of mutual drag between phonons and spin excitations
on the thermal conductivity of a quantum spin system is discussed.
We derive general expression for the drag component of the
thermal current using Boltzmann equation as well as Kubo
linear-response formalism to leading order in the
spin-phonon coupling. 
We demonstrate that aside from
higher-order corrections which appear in the Kubo formalism
both approaches
yield identical result for the drag thermal conductivity.
We discuss the range of applicability of our result and provide
a generalization of our consideration to the cases of fermionic
excitations and to anomalous forms of boson-phonon coupling.
Several asymptotic regimes of our findings relevant to realistic
situations are highlighted.
\end{abstract}
\maketitle

\section{Introduction}
\label{intro}

Transport phenomena form a prominent group of problems in condensed matter physics.
They provide a unique information on the excitations and their interactions not accessible by other methods.\cite{Ziman,Abrikos} Recently, \textit{thermal} transport by spin excitations in low-dimensional quantum magnets has received significant attention due to very large heat conductivities found in a number of materials, for reviews see Refs. \onlinecite{Hess2007}, \onlinecite{fhm07}.  
One may speculate that thermal conductivity could be used to probe
elementary excitations in quantum magnets in a fashion analogous to the use of
electrical conductivity in metals.

By now it is well established that integrable one-dimensional quantum magnets
allow for infinite heat conductivity.\cite{Zotos1997,Prelovsek,fhm07}
Experimentally, however, many spin systems rather remote from integrability also
demonstrate large heat conductivities.\cite{Hess2001,Hess2007}
 Understanding the role of
coupling of the spin degrees of freedom to an environment, such as phonons and
impurities, could be essential in this context.
Phonons are ubiquitous heat
carriers along with spin excitations in all quantum magnets. Usually,
interaction between spins and phonons is discussed in the context
of dissipation of their respective currents.
Significant progress has been made here,\cite{Shimshoni2003,Rozhkov2005,Louis2006} yet, many questions remain open.

In this work we focus on one such question which is rarely addressed: the off-diagonal effect of the flow of one of the excitations
facilitating the flow of the other.\cite{Gurevich_67,Boulat2007,Chernyshev07}
It is referred to as ``spin-phonon drag'', in analogy with electron-phonon drag
discussed in the thermoelectric phenomena in metals and
semiconductors\cite{Berman,mahan1990,thermpwr,Holstein1964,Bass90,Ziman_eph,Gurevich46}.

The second question we address in this work is the relation between two distinct
theoretical approaches to transport in a generic coupled two-component system,
namely the quasi-classical Boltzmann transport theory and the
Kubo linear-response
formalism. Such relations, while of fundamental importance, remain
unclear between many techniques\cite{Rammer1986,Mori1965,Goetze1972,Langer1960,Konstantinov1960,Yamada1962,Michel1964a,Michel1964b,Ron1964,Plakida1964} devised in the past. For selected problems and techniques such correspondence has been established rather firmly,\cite{Hanch_Mahan_83,mahan1990} but, to the best of our knowledge, the comparison discussed in this work has not been
performed previously.

Historically, the term phonon drag appears in two rather distinct
contexts. The first connotation is the negative effect of
phonons on the electrical conductivity by slowing down electrons via processes
that are different from the direct scattering effects.\cite{Holstein1964,Bass90}
The second, also referred to as the Gurevich effect,\cite{Ziman,Gurevich46} is
responsible for the dramatic deviation of the thermopower in many materials from
the predictions of the ``electron-only'' theory.\cite{Ziman_eph} In
its idealized version,\cite{Ziman,Gurevich46} the phonon drag results in a
substantial heat flow due to adjustment of the momentum distribution of phonons
to that of the electrons, the latter being displaced by the electric field.

In this work, the ``thermal-only'' analog of the Gurevich effect for a generic
spin-phonon problem is considered. In this case, only the thermal current is of
interest. In the standard electron-phonon problem, the thermal-only drag is
discarded traditionally. This is because the thermal conductivity by phonons in
metals can usually be neglected due to strong scattering of phonons and the very
large ratio of the Fermi to sound velocity.\cite{Ziman} However, this is not
the case in several magnetic insulators of current interest
\cite{Hess2001,Sologubenko2001,Sologubenko2003,Ribeiro2005,Hess2007,Hess2010} where the magnetic and
lattice heat conductivity can be of the same order. Therefore, the drag between
spin excitations and phonons can be an important phenomenon. We also note in
passing, that in contrast the previous electron-phonon problems the
dimensionality of the spin and the phonon system in magnetic insulators can be
different. Spin excitations may be confined to chains, ladders, and planes.
Thus, the focus of our study is on the general problem of a two-component system
and the drag effect in the thermal conductivity.

While the spin-phonon problem is our main motivation,
we consider a generic model of bosonic quasiparticle-like excitations,
e.g. magnons, coupled to phonons. We derive the contribution
of the phonon drag to the thermal conductivity in the lowest order of
the boson-phonon coupling using Kubo and Boltzmann formalisms.
We demonstrate that both approaches yield \textit{identical} results
for the drag component of the thermal conductivity, thus establishing
a direct correspondence between these methods. We note that despite
a significant body of work on thermoelectric phenomena, to the best
of our knowledge, such results have not been discussed before within
these two approaches.

While most of the work is devoted to bosonic spin excitations, we have also
generalized our consideration to the case of fermionic
excitations, as well as to the case of particle
non-conserving boson-phonon interactions. The latter are as common as the
``normal'', particle-conserving ones and occur, e.g., in the
phases with broken symmetry. We also note that using
our expressions for the drag thermal conductivity we reproduce some
of the results of the recent work,\cite{Boulat2007}
which considers spin-phonon drag in a particular class of
quantum magnets using the memory-function approach.

The paper is organized as follows. In Sec. \ref{sec:Model} we introduce the
model. Section \ref{Boltzmann} outlines the derivation of the drag conductivity
from the Boltzmann equation. In Section \ref{Kubo} we detail the Kubo diagrams
for drag and elaborate on one approach to their evaluation. Sec. \ref{other}
extends the consideration of the drag thermal conductivity onto fermionic
excitations as well as onto anomalous (particle-non-conserving) boson-phonon
coupling.  In Sec. \ref{sec:qualitative} we provide a  qualitative
discussion of thermal drag for several representative cases and asymptotic
regimes. We conclude with Sec. \ref{sec:Conclusion}.  Several Appendices are
provided. In Appendix \ref{sec:DetailsBZA} we detail more technical points of
the Boltzmann approach. Appendix \ref{sec:Analytic-continuation} is devoted to
the evaluation of the Kubo diagrams for the drag using an alternative approach.
Finally, in Appendix \ref{sec:NBC} we discuss some corrections beyond the
Boltzmann results from the Kubo diagrams.

\section{Model}
\label{sec:Model}

Spin systems can yield a wide variety of excitations, such as spinons, magnons,
triplet excitations, etc., depending on the dimensionality, spin value, and
geometry of the structural arrangement.  Because the focus of this work is on
the drag between spin excitations and phonons, we assume that both of
them are describable by well-defined quasiparticles. For the spin
excitations we assume energy dispersion $\varepsilon_{\bf k}$ and some
phenomenological intrinsic or extrinsic transport relaxation rate. This rate may
include scattering caused by impurities, spin-spin interaction, and other
degrees of freedom.  The corresponding energy for phonons will be denoted by
$\omega_{\bf p}$ and they are also assumed to have a finite relaxation rate.  In most of the paper, we assume the statistics of the
spin excitations to be that of bosons. Sec.~\ref{other} outlines the changes in
the drag conductivity which results if the choice of the statistics would be
that of fermions.

The spin-phonon coupling occurs because lattice
displacements cause changes in the spin interactions and
anisotropies.\cite{Dixon} Thus, the simplest, yet very
general form of the spin-phonon coupling is {\it linear} 
in the lattice displacement and quadratic in the spin operators.
After mapping spins onto bosonic quasiparticles, the resultant
lowest-order boson-phonon coupling will generally
contain the ``normal'' part, which conserves the boson number,\cite{Rozhkov2005} and the off-diagonal, ``anomalous'' bosonic terms.\cite{Dixon}
Since the subsequent drag conductivity derivation is conceptually identical for the normal and anomalous forms of interactions, we will postpone
consideration of the latter until Sec.~\ref{other} and will treat it in less detail. Here and in the next two Sections, we will concentrate on the normal form of the coupling.

Altogether, the Hamiltonian implied for our subsequent consideration is
\begin{eqnarray}
\label{H}
\vphantom{\sum_{\mathbf{k}}}
&&H = H_{\mathrm{b}}+
H_{\mathrm{ph}}+H_{\mathrm{b-ph}}\,,\label{eq:-1}\\
\label{H0}
&&H_{\mathrm{b}}=\sum_{\mathbf{k}}\varepsilon_{\mathbf{k}}b_{\mathbf{k}}^{\dagger}
b_{\mathbf{k}}^{\phantom{\dagger}}\,,\ \ \ \
H_{\mathrm{ph}} =  \sum_{\mathbf{p}}\omega_{\mathbf{p}}a_{\mathbf{p}}^{\dagger}
a_{\mathbf{p}}^{\phantom{\dagger}}\,,\label{eq:-3}\\
&&H_{\mathrm{b-ph}}=  \sum_{\mathbf{k},\mathbf{p}}
V_{\mathbf{p};\mathbf{k},\mathbf{k}-\mathbf{p}}^{\rm b-ph}
b_{\mathbf{k}-\mathbf{p}}^{\dagger}b_{\mathbf{k}}^{\phantom{\dagger}}
(a_{\mathbf{p}}^{\dagger}+a_{-\mathbf{p}}^{\phantom{\dagger}})\,,
\hphantom{aaaaaaaaa}
\label{Hint}
\end{eqnarray}
where $b_{\mathbf{k}}^{(\dagger)}$ and $a_{\mathbf{p}}^{(\dagger)}$
are boson and phonon operators, and
$V_{-\mathbf{p};\mathbf{k}-\mathbf{p},\mathbf{k}}^{\rm b-ph}
=(V_{\mathbf{p};\mathbf{k},\mathbf{k}-\mathbf{p}}^{\rm b-ph})^{\star}$
due to hermiticity of $H_{\rm b-ph}$, and we do not
specify interaction terms that result in the
relaxation rates of bosons and phonons.
Note that, aside from a
more general momentum dependence,
the boson-phonon interaction in (\ref{Hint}) is the
same as the one in the electron-phonon coupling case.

\section{Boltzmann approach}
\label{Boltzmann}
\subsection{Thermal drag conductivity}

Let us denote boson and phonon distribution functions as
$f_{\bf k}$ and $n_{\bf p}$, respectively.
In Boltzmann's approach the total heat current is the sum of the
currents from each of the particle species:
\begin{eqnarray}
\label{j0}
{\bf j}_{\rm tot}=\sum_{\bf k} {\bf v}_{\bf k}
\varepsilon_{\bf k} \delta f_{\bf k}+\sum_{\bf p} {\bf u}_{\bf p}
\omega_{\bf p} \delta n_{\bf p},
\end{eqnarray}
where ${\bf v}_{\bf k}=\partial\varepsilon_{\bf k}/\partial{\bf k}$ and
${\bf u}_{\bf p}=\partial\omega_{\bf p}/\partial{\bf p}$ are the velocities,
the chemical potential is set to
zero, and $\delta f_{\bf k}$ and $\delta n_{\bf p}$ are
the non-equilibrium parts of the distribution functions of the bosons and phonons, respectively.

The distribution functions are determined from the Boltzmann equations
\begin{eqnarray}
\label{BE0}
\frac{d f_{\bf k}}{d t}= St^{\rm b}_{\bf k}[f,n]\, ,\ \ \
\frac{d n_{\bf p}}{d t}= St^{\rm ph}_{\bf p}[n,f]\,,
\end{eqnarray}
where $St^{\rm b (ph)}$ are the collision integrals which include all possible
scatterings for bosons (phonons). These Boltzmann equations are coupled because
the boson-phonon interaction in (\ref{Hint}) yields
terms in the collision integrals which depend on both $f_{\bf k}$ and
$n_{\bf p}$.

Assuming the system to be in a steady state
under a small uniform thermal gradient, we may linearize the Boltzmann
equations in  $\delta f$ and $\delta n$ to find
\begin{eqnarray}
\label{BE0a}
&&
\frac{\varepsilon_{\bf k}}{T}\,
\frac{\partial f^0_{\bf k}}{\partial \varepsilon}
\left({\bf v}_{\bf k}\cdot\mbox{\boldmath$\nabla$}T\right)=
-\frac{\delta f_{\bf k}}{\tau^{\rm b}_{\bf k}}
-\sum_{\bf p}\frac{\delta n_{\bf p}}{\tau^{\rm ph\rightarrow b}_{\bf p,k}}
,\\
\label{BE0b}
&&
\frac{\omega_{\bf p}}{T}\,
\frac{\partial n^0_{\bf p}}{\partial \omega}
\left({\bf u}_{\bf p}\cdot\mbox{\boldmath$\nabla$}T\right)=
-\frac{\delta n_{\bf p}}{\tau^{\rm ph}_{\bf p}}
-\sum_{\bf k}\frac{\delta f_{\bf k}}{\tau^{\rm b\rightarrow ph}_{\bf k,p}},
\end{eqnarray}
where $f_{\bf k}^0=[e^{\varepsilon_{\bf k}/T}-1]^{-1}$ and
$n_{\bf p}^0=[e^{\omega_{\bf p}/T}-1]^{-1}$ are the equilibrium distribution functions
and the collision integrals in the right-hand sides are expanded in $\delta f$
and $\delta n$ and are considered
in the relaxation-time approximation. The first
terms on the right-hand sides of (\ref{BE0a}) and (\ref{BE0b}) are the usual
diffusion terms, with $1/\tau^{\rm b} _{\bf k}$ and $1/\tau^{\rm ph}_{\bf p}$
being the transport relaxation rates of the bosons and phonons due to all possible relaxation mechanisms, as discussed in
Sec.~\ref{sec:Model}.  The second term on the right-hand side of the
Boltzmann equation (\ref{BE0a}) for bosons with the momentum ${\bf k}$ is
from the expansion of the collision integral $St^{\rm b}_{\bf k}[f,n]$ in the
phonon non-equilibrium distribution $\delta n_{\bf p}$. Because of that it  contains an integral over the phonon momentum.
The same is true for the phonon Boltzmann
equation (\ref{BE0b}).  These latter terms arise solely due to the
boson-phonon coupling (\ref{Hint}) and are due to the non-equilibrium
components of the particles of opposite species.  Therefore, it is natural to
identify them with the drag from one species of excitations onto the other.
Below we are going to explicate the relation of $1/\tau^{\rm ph\rightarrow
b}_{\bf p,k}$ and $1/\tau^{\rm b\rightarrow ph}_{\bf k,p}$ with
$V_{\mathbf{p};\mathbf{k},\mathbf{k}-\mathbf{p}}^{\rm b-ph}$ through the
boson-phonon collision integral, but at this stage we simply use them as a
short-hand notations for the ``drag rates''.\cite{Herring}

In general, Eqs. (\ref{BE0a}) and (\ref{BE0b}) reduce to integral
equations for $\delta f$ and $\delta n$. However, we assume that the drag
terms in (\ref{BE0a}) and (\ref{BE0b}) are small compared to the diffusion
contribution, i.e., the drag rates
$1/\tau^{\rm ph\rightarrow b}$ and $1/\tau^{\rm b\rightarrow ph}$
are small compared to the intrinsic boson and phonon rates
$1/\tau^{\rm b}$ and $1/\tau^{\rm ph}$.
This is equivalent to treating the boson-phonon coupling $V^{\rm b-ph}$ as a perturbation.
In turn, one can solve  (\ref{BE0a}) and (\ref{BE0b})
iteratively by using the diffusion-only components in the integrals containing
$\delta n$ and $\delta f$. This corresponds to neglecting the terms of order
$|V^{\rm b-ph}|^4$ and higher.  Within this approximation,
the total current in (\ref{j0}) is
\begin{eqnarray}
\label{j1}
{\bf j}_{\rm tot}={\bf j}_{\rm b} + {\bf j}_{\rm ph}+{\bf j}_{\rm d, ph\rightarrow b} +{\bf j}_{\rm d, b\rightarrow ph},
\end{eqnarray}
where ${\bf j}_{\rm b}$ and ${\bf j}_{\rm ph}$ are the usual, ``diagonal''
terms, and ${\bf j}_{\rm d, ph\rightarrow b}$ and ${\bf j}_{\rm d, b\rightarrow
ph}$ are the currents due to the drag of phonons on bosons and vice versa.
The total drag current can be written as:
\begin{eqnarray}
\label{j3}
j^\alpha_{\rm drag}&=&j^\alpha_{\rm d, ph\rightarrow b}+j^\alpha_{\rm d, b\rightarrow ph}
=-\frac{1}{T}\sum_{\bf k,p} \tau^{\rm b}_{\bf k}\,\tau^{\rm ph}_{\bf p}\,
\varepsilon_{\bf k}\,\omega_{\bf p}
\nonumber\\
&&\times\bigg[\frac{\partial n^0_{\bf p}}{\partial \omega}\,
\frac{v^\alpha_{\bf k}\,u^\beta_{\bf p}}{\tau^{\rm ph\rightarrow b}_{\bf p,k}}
+\frac{\partial f^0_{\bf k}}{\partial \varepsilon}\,
\frac{u^\alpha_{\bf p}\,v^\beta_{\bf k}}{\tau^{\rm b\rightarrow ph}_{\bf k,p}}\bigg]
\, \nabla_\beta T,
\end{eqnarray}
where $\alpha$ and $\beta$ are vector components.
Choosing the temperature gradient in the $x$-direction and assuming
the conductivity tensor to be diagonal, we obtain the
drag thermal conductivity:
\begin{eqnarray}
\label{kappa}
\kappa_{\rm drag}&=&\frac{1}{T}\sum_{\bf k,p}
\left(v^x_{\bf k}\,\varepsilon_{\bf k}\,\tau^{\rm b} _{\bf k}\right)\,
\left(u^x_{\bf p}\,\omega_{\bf p}\,\tau^{\rm ph}_{\bf p}\right)\,\nonumber\\
&&\phantom{-\sum_{\bf k,p}}
\times\bigg[\frac{\partial n^0_{\bf p}}{\partial \omega}\,
\frac{1}{\tau^{\rm ph\rightarrow b}_{\bf p,k}}+\frac{\partial f^0_{\bf k}}{\partial \varepsilon}\,
\frac{1}{\tau^{\rm b\rightarrow ph}_{\bf k,p}}\bigg],
\end{eqnarray}
The above analysis thus far has been independent of the microscopic form
of the boson-phonon coupling.

\subsection{Microscopic consideration.}

 The ``drag rates'' $1/\tau^{\rm ph\rightarrow b}_{\bf p,k}$ and $1/\tau^{\rm
b\rightarrow ph}_{\bf k,p}$ are obtained by taking variations of $f$ and $n$ in
the corresponding functionals $St[f,n]$. We now detail the derivation of one of
them.  The collision integral for phonons scattered off bosons via
Eq.~(\ref{Hint}) contains two terms: the first one increases the number of
phonons with momentum ${\bf p}$, the second one reduces it.  They can be grouped
together as:
\begin{eqnarray}
\label{Stn}
&&St^{\rm ph}_{\bf p}[n,f]=2\pi\sum_{\bf k}
\big|V^{\rm b-ph}_{{\bf p};{\bf k},{\bf k}-{\bf p}}\big|^2
\cdot\delta\big(\varepsilon_{\bf k}-
\varepsilon_{\bf k-p}-\omega_{\bf p}\big)
\nonumber\\
&&\phantom{St^{\rm ph}_{\bf p}[n,f]=}
\times\big[f_{\bf k} (f_{\bf k-p}+n_{\bf p}+1)
-f_{\bf k-p} n_{\bf p}\big]
\end{eqnarray}
This is the complete expression of the phonon collision integral
due to phonon-boson interaction (\ref{Hint}). The subsequent
linearization of (\ref{Stn}) uses the condition
$St^{\rm ph}_{\bf p}[f^0,n^0]\equiv 0$. Writing $f=f^0+\delta f$ and
$n=n^0+\delta n$ and neglecting terms of order $\delta f\delta n$ and
$(\delta f)^2$ yields
the first and second terms in the right-hand side of Eq.~(\ref{BE0b}). The
first one is the diffusion  term while
the second one is the drag term. In the latter, for the terms containing
$\delta f_{\bf k-p}$, we shift summation over
 ${\bf k}\rightarrow {\bf k}+{\bf p}$
so that $\delta f_{\bf k-p}\rightarrow\delta f_{\bf k}$.
After these manipulations,
the drag rate of bosons on phonons is given by
\begin{eqnarray}
\label{taufn}
&&St^{\rm ph}_{\bf p}[n^0,f^0+\delta f]\approx
-\sum_{\bf k}\frac{\delta f_{\bf k}}{\tau^{\rm b-ph}_{\bf k,p}}
,\nonumber\\
&&\frac{1}{\tau^{\rm b\rightarrow ph}_{\bf k,p}}=-2\pi
\Big(\big|V^{\rm b-ph}_{{\bf p};{\bf k},{\bf k}-{\bf p}}\big|^2
\big[f^0_{\bf k-p} +n^0_{\bf p}+1\big]
\\
&&\phantom{\frac{1}{\tau^{\rm ph-b}_{\bf p,k}}=-2\pi\Big(\ \ \ \ \ \ \ \ }
\times\delta\big(\varepsilon_{\bf k}-
\varepsilon_{\bf k-p}-\omega_{\bf p}\big)\nonumber\\
&&
+ \big|V^{\rm b-ph}_{{\bf p};{\bf k}+{\bf p},{\bf k}}\big|^2 \big[f^0_{\bf k+p} - n^0_{\bf p}\big]
\cdot\delta\big(\varepsilon_{\bf k}-
\varepsilon_{\bf k+p}+\omega_{\bf p}\big)\Big).\nonumber
\end{eqnarray}
Note that under  ${\bf p}\rightarrow -{\bf p}$, the second  term in (\ref{taufn})  changes  to $\big|V^{\rm b-ph}_{{\bf -p};{\bf k}-{\bf p},{\bf k}}\big|^2= \big|V^{\rm b-ph}_{{\bf p};{\bf k},{\bf k}-{\bf p}}\big|^2$ and the phonon velocity changes its sign, ${\bf u}_{\bf p}\rightarrow -{\bf u}_{\bf -p}$, see discussion after (\ref{Hint}). Using this symmetry we obtain a compact form for the component of the thermal conductivity due to the drag of bosons on phonons:
\begin{eqnarray}
\label{kappa2phb1}
&&\kappa_{\rm drag}^{\rm b\rightarrow ph}=-\frac{2\pi}{T}\sum_{\bf k,p}
\left(v^x_{\bf k}\,\varepsilon_{\bf k}\,\tau^{\rm b} _{\bf k}\right)\,
\left(u^x_{\bf p}\,\omega_{\bf p}\,\tau^{\rm ph}_{\bf p}\right)\,
\big|V^{\rm b-ph}_{{\bf p};{\bf k},{\bf k}-{\bf p}}\big|^2
\nonumber\\
&&\phantom{\kappa_{\rm drag}}
\times\bigg[\frac{\partial f^0_{\bf k}}{\partial \varepsilon}\,
\big[f^0_{\bf k-p} +n^0_{\bf p}+1\big]
\delta\big(\varepsilon_{\bf k}-
\varepsilon_{\bf k-p}-\omega_{\bf p}\big)\ \ \ \ \ \\
&&\phantom{\kappa_{\rm drag}=}
-\frac{\partial f^0_{\bf k}}{\partial \varepsilon}\,
\big[f^0_{\bf k-p} -n^0_{\bf p}\big]
\delta\big(\varepsilon_{\bf k}-
\varepsilon_{\bf k-p}+\omega_{\bf p}\big)
\bigg].\nonumber
\end{eqnarray}
The derivation of the drag rate $1/\tau^{\rm ph\rightarrow b}_{\bf p,k}$
and of the drag conductivity of phonon on
bosons follows similar reasoning and is presented in Appendix \ref{app_A}.1.
After some algebra one arrives at the following statement:
\begin{eqnarray}
\label{kappa2equal}
\kappa_{\rm drag}^{\rm b\rightarrow ph}\equiv
\kappa_{\rm drag}^{\rm ph\rightarrow b}.
\end{eqnarray}
This relation bears a simple and general physical meaning: within linear
response, the non-equilibrium component of one species of particles causes the
same drag on the  other species as the non-equilibrium component of the
other causes on the first one.  That is, phonons drag bosons the same as
bosons drag phonons. Thus, the total contribution to the conductivity is simply
twice the contribution in Eq.~(\ref{kappa2phb1}).
In addition to the algebra above, to obtain
(\ref{kappa2equal}) we have used the following identities between the
combinations of the bosonic distribution functions and their derivatives,
\begin{eqnarray}
\label{convert1}
\frac{\partial n^0_{\bf p}}{\partial \omega}
\big[f^0_{\bf k-p} -f^0_{\bf k}\big]\equiv
\frac{\partial f^0_{\bf k}}{\partial \varepsilon}
\big[f^0_{\bf k-p} +n^0_{\bf p}+1\big]\Bigg|_{\varepsilon_{\bf k}-
\varepsilon_{\bf k-p}=\omega_{\bf p}}\\
\label{convert2}
\frac{\partial n^0_{\bf p}}{\partial \omega}
\big[f^0_{\bf k-p} -f^0_{\bf k}\big]\equiv
\frac{\partial f^0_{\bf k}}{\partial \varepsilon}
\big[f^0_{\bf k-p} -n^0_{\bf p}\big]\Bigg|_{\varepsilon_{\bf k-p}-
\varepsilon_{\bf k}=\omega_{\bf p}}\, ,
\end{eqnarray}
which can be obtained with the help of $St^{\rm ph}_{\bf p}[f^0,n^0]\equiv 0$.

Thus, within the Boltzmann formalism, the total drag thermal conductivity,
to leading order in the boson-phonon coupling, is given by
\begin{eqnarray}
\label{kappatotal}
\kappa_{\rm drag} &=& -\frac{4\pi}{T}\sum_{\bf k,p}
\left(v^x_{\bf k}\,\varepsilon_{\bf k}\,\tau^{\rm b} _{\bf k}\right)\,
\left(u^x_{\bf p}\,\omega_{\bf p}\,\tau^{\rm ph}_{\bf p}\right)\,
\big|V^{\rm b-ph}_{{\bf p};{\bf k},{\bf k}-{\bf p}}\big|^2
\nonumber\\
&&
\times\bigg[\frac{\partial f^0_{\bf k}}{\partial \varepsilon}\,
\big[f^0_{\bf k-p} +n^0_{\bf p}+1\big]
\delta\big(\varepsilon_{\bf k}-
\varepsilon_{\bf k-p}-\omega_{\bf p}\big)\ \ \ \ \ \\
&&
-\frac{\partial f^0_{\bf k}}{\partial \varepsilon}\,
\big[f^0_{\bf k-p} -n^0_{\bf p}\big]
\delta\big(\varepsilon_{\bf k}-
\varepsilon_{\bf k-p}+\omega_{\bf p}\big)
\bigg].\nonumber
\end{eqnarray}
This expression is the main result of this work.
We would like to note, that within the approximations discussed above this
result is valid for any type of scattering, impurity, boundary or Umklapp, all
being implicitly incorporated in the transport relaxation times of phonons and
bosons. This expression also contains normal as well as the Umklapp boson-phonon
scattering. That is, the quasimomenta in (\ref{kappatotal}) are defined up to
the reciprocal lattice vectors and the summation over the latter is assumed as
usual.\cite{LLX}

\section{Kubo Approach}
\label{Kubo}

A different theoretical approach to transport, alternative to the Boltzmann equation, is the Kubo linear-response formalism. The great advantage of this
approach is its conceptual clarity with regard to the definition of the drag
thermal conductivity.  It is also very effective in classifying terms by
their respective order in the coupling constant as it contains them explicitly.

In Kubo's approach the uniform part of the thermal conductivity is obtained
by taking the DC-limit of the imaginary part of the dynamical heat-current
susceptibility $\chi^{\mu\nu}$:\cite{mahan1990}
\begin{eqnarray}
\kappa^{\mu\nu}=-\lim_{\omega\rightarrow0}\frac{\beta}{\omega+i0}\ {\rm Im}
\left[\chi^{\mu\nu}\left(0,\omega+i0\right)\right]\,,\label{eq:2}
\end{eqnarray}
where $\mu$ and $\nu$ are the spatial directions and the susceptibility is a sum
of diagonal and off-diagonal terms,
\begin{eqnarray}
\chi^{\mu\nu}=\sum_{i,j=1,2} \chi^{\mu\nu}_{i,j}\, ,
\label{eq:2a}
\end{eqnarray}
with the components
\begin{eqnarray}
\label{eq:2b}
\chi^{\mu\nu}_{i,j}\left(\mathbf{q},\omega+i0\right)=i\int_{0}^{\infty}\langle
[j_{\mathbf{q} i}^{\mu}(t),j_{-\mathbf{q} j}^{\nu}]\rangle e^{i(\omega+i0)t}dt ,
\end{eqnarray}
which contain the heat-currents $j_{\mathbf{q} i}^{\mu}$. In this study,
the long wavelength limit of the thermal current of bosons is given by
\begin{eqnarray}
\mathbf{j}_{\mathbf{q} 1}&=&\sum_{\mathbf{k}}
\varepsilon_{\mathbf{k}}v_{\mathbf{k}} b_{\mathbf{k}+\mathbf{q}}^{\dagger}b_{\mathbf{k}}^{\phantom{\dagger}},\label{spin-boson}
\end{eqnarray}
and the phonon one by
\begin{eqnarray}
\mathbf{j}_{\mathbf{q} 2}&=&\sum_{\mathbf{p}}
\omega_{\mathbf{p}}u_{\mathbf{p}} a_{\mathbf{p}+\mathbf{q}}^{\dagger}a_{\mathbf{p}}^{\phantom{\dagger}},\label{phonon}
\end{eqnarray}
where the energies and velocities were defined previously in (\ref{H0}) and
(\ref{j0}). For the remainder of the paper the usual limit of ${\bf q}=0$ is
implied for the currents, however ${\bf q}$ is kept visible for clarity. Note
that Eq.~(\ref{eq:2}) is derived from the linear response to the temperature
gradient, which couples to the {\em total} energy density.
Therefore, apart from the bare
heat currents of (\ref{spin-boson}) and (\ref{phonon}), the interaction
term in the Hamiltonian (\ref{Hint}) will also
give rise to a contribution to the thermal current.
This current, labeled by $\mathbf{j}_{3,\mathbf{q}}$, follows from the continuity equation,
\begin{eqnarray}
\mathbf{q}\cdot\mathbf{j}_{3,\mathbf{q}}=[H,H_{\mathbf{q}}]-
\mathbf{q}\cdot(\mathbf{j}_{1,\mathbf{q}}+ \mathbf{j}_{2,\mathbf{q}}),\label{continuity}
\end{eqnarray}
where $H_{\mathbf{q}} = \sum_{\bf r} e^{-i{\bf q}\cdot{\bf r}} H_{\bf r}$ is the
Fourier transform of a position-dependent Hamiltonian energy density, $H =
\sum_{\bf r} H_\mathbf{r}$. However, $\mathbf{j}_{3,\mathbf{q}}$ does not
constitute a contribution to thermal drag and we will not consider the
corresponding terms in this work.

\begin{figure}[t]
\centering
\includegraphics[width=0.99\columnwidth]{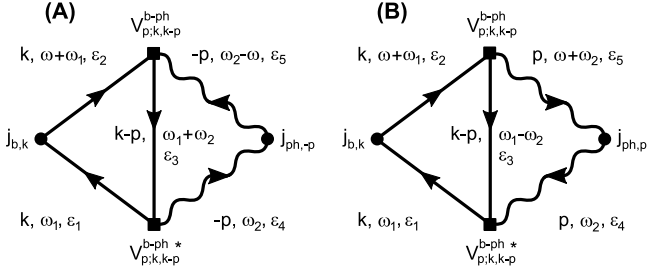}
\caption{Graphical representation of the lowest order, off-diagonal
current-current correlations contributing to boson-phonon drag
conductivity. Solid lines are bosons, wavy lines are phonons.
$\varepsilon$'s are the auxiliary frequencies used in the
spectral representation approach.}
\label{diagrams}
\end{figure}

We would like
to emphasize that (\ref{eq:2}) refers only to the DC-limit and does \emph{not}
incorporate the Drude weight.\cite{fhm07} The latter is assumed to be zero
henceforth.

The diagonal ($i\!=\!j$) components in (\ref{eq:2b}) are the ``usual''
diffusion terms and they do not contribute to the drag.
Naturally, the drag is given by the off-diagonal current-current correlation functions, $\chi_{1,2}$ and $\chi_{2,1}$.
Considering the boson-phonon coupling $V_{\mathbf{p};\mathbf{k},\mathbf{k}-\mathbf{p}}^{\rm b-ph}$ in (\ref{Hint})
as a perturbation, the lowest-order diagrams contributing to the drag are shown in Fig.~\ref{diagrams}. These two diagrams are the only ``drag'' diagrams of the order $|V^{\rm b-ph}|^2$ that contribute to $\chi_{1,2}$. The mirror-reflection of these diagrams with respect to a vertical line yields equivalent contributions to $\chi_{2,1}$. This is, again, a
graphical way of stating that in the linear response the drag of
phonons on bosons and the one from bosons on phonons are identical.
Thus, the total drag conductivity is given by twice the value of the diagrams in Fig.~\ref{diagrams}.
The choice of the momenta in Fig.~\ref{diagrams}  is made to keep all the
momentum-dependent functions between the two diagrams, such as energies and vertices, the same. Since in the diagram $A$ the phonon
momentum is ${\bf -p}$, its sign is opposite
due to the velocity in the current vertex.
The analytical expression for the total drag thermal
conductivity given by Fig.~\ref{diagrams} and its mirror-reflection is:
\begin{eqnarray}
\label{K_kappa0}
\kappa_{\rm drag}=\lim_{\omega\rightarrow 0}\bigg[
\frac{2}{T\omega}\sum_{\bf k,p}
\left(v^x_{\bf k}\,\varepsilon_{\bf k}\right)\,
\left(u^x_{\bf p}\,\omega_{\bf p}\right)\,
\big|V^{\rm b-ph}_{{\bf p};{\bf k},{\bf k}-{\bf p}}\big|^2\nonumber\\
\phantom{\kappa_{\rm drag} }
{\rm Im}\Big(\Pi_A({\bf k},{\bf p},\omega+i0)
-\Pi_B({\bf k},{\bf p},\omega+i0)\Big)\bigg],
\end{eqnarray}
with
\begin{eqnarray}
\label{IA}
&&\Pi_{A}({\bf k},{\bf p},i\omega)=T^2\sum_{\omega_1,\omega_2}
G_{\bf k}(i\omega+i\omega_1)G_{\bf k}(i\omega_1)\\
&&\phantom{\Pi_{A}({\bf k},{\bf p},i\omega)}
\times G_{\bf k-p}(i\omega_1+ i\omega_2){\bar G}_{\bf p}(i\omega_2- i\omega)
{\bar G}_{\bf p}(i\omega_2)
,\nonumber
\end{eqnarray}
and
\begin{eqnarray}
\label{IB}
&&\Pi_{B}({\bf k},{\bf p},i\omega)=
T^2\sum_{\omega_1,\omega_2}
G_{\bf k}(i\omega+i\omega_1)G_{\bf k}(i\omega_1)\\
&&\phantom{\Pi_{A}({\bf k},{\bf p},i\omega)}
\times G_{\bf k-p}(i\omega_1- i\omega_2){\bar G}_{\bf p}(i\omega_2+ i\omega)
{\bar G}_{\bf p}(i\omega_2)
,\nonumber
\end{eqnarray}
where $G_{\bf k}(i\omega)$ and ${\bar G}_{\bf p}(i\omega)$
are the boson and phonon Green's functions, respectively.

To calculate the thermal conductivity one needs to perform the frequency
summations in (\ref{IA}) and (\ref{IB}). We utilize two technical approaches
for that: the first uses the spectral representation for
the Matsubara Green's functions and the second one uses integration along the
branch cuts of the Green's functions.\cite{mahan1990}
Below we elaborate on the use of the
first one while the branch cut integration approach
is discussed in Appendix \ref{sec:BCA}.

\subsection{Spectral Representation Approach\label{sub:Spectral-Analysis}}

The spectral representation for the Matsubara Green's function is:
\begin{eqnarray}
\label{rep}
G_{\bf k}(i\omega_n)=\int_\varepsilon\,
\frac{A_{\bf k}(\varepsilon)}{i\omega_n-\varepsilon}\ ,
\end{eqnarray}
where the shorthand notation
$\int_\varepsilon\equiv\int_{-\infty}^\infty
d\varepsilon/2\pi$ and
the following relation of the spectral function
to the retarded Green's function, $A_{\bf k}(\varepsilon)=-2\,{\rm Im}\,
 G_{\bf k}^R(\varepsilon)$, are used.

Below we calculate the contributions to the thermal conductivity
from the diagram $A$ in Fig.~\ref{diagrams}, those from the diagram $B$
are discussed near the end of the section.
Using the spectral function representation (\ref{rep}),
 and assigning auxiliary
frequencies according to the diagrams in Fig.~\ref{diagrams},
one can rewrite $\Pi_{A}$ as:
\begin{eqnarray}
\label{IAB_1}
\Pi_A({\bf k},{\bf p},i\omega)&=&\int_{\varepsilon_1,\dots\varepsilon_5}
A_{\bf k}(\varepsilon_1)A_{\bf k}(\varepsilon_2)
A_{\bf k-p}(\varepsilon_3)\\
&&\times {\bar A}_{\bf p}(\varepsilon_4)
{\bar A}_{\bf p}(\varepsilon_5)\cdot
\overline{\Pi}_{A}(i\omega,\varepsilon_{1\dots 5}),\nonumber
\end{eqnarray}
where $\varepsilon_{1\dots 5}$ stands for the five frequencies
associated with each individual line in Fig.~\ref{diagrams}(A).
The frequency summation over $\omega_1$ and $\omega_2$
is now accumulated in $\overline{\Pi}_{A}$, which is given by:
\begin{eqnarray}
\label{IAB_2}
&&\overline{\Pi}_{A}(i\omega,\varepsilon_{1\dots 5})=
T^2\sum_{\omega_1,\omega_2}
\frac{1}{i\omega_1-\varepsilon_1}\cdot
\frac{1}{i\omega_1+i\omega-\varepsilon_2}\nonumber\\
&&\phantom{{\bar I}}
\cdot
\frac{1}{i\omega_1+ i\omega_2-\varepsilon_3}\cdot
\frac{1}{i\omega_2-\varepsilon_4}\cdot
\frac{1}{i\omega_2- i\omega-\varepsilon_5}\ .
\end{eqnarray}
Performing Matsubara frequency summations in (\ref{IAB_2})
we obtain:
\begin{eqnarray}
\label{IAB_3a}
&&\overline{\Pi}_{A}(i\omega,\varepsilon_{1\dots 5})=
\frac{1}{i\omega+\varepsilon_1-\varepsilon_2}\cdot
\frac{1}{i\omega+\varepsilon_5-\varepsilon_4}\\
&&\phantom{{\bar I}_{A}}
\bigg[
\frac{(n_3-n_1)(n_{3-1}-n_5)}
{i\omega +\varepsilon_5+\varepsilon_1-\varepsilon_3}-
\frac{(n_3-n_1)(n_{3-1}-n_4)}
{\varepsilon_4+\varepsilon_1-\varepsilon_3}\nonumber\\
&&\phantom{{\bar I}_{A}}
+\frac{(n_3-n_2)(n_{3-2}-n_4)}
{-i\omega +\varepsilon_4+\varepsilon_2-\varepsilon_3}-
\frac{(n_3-n_2)(n_{3-2}-n_5)}
{\varepsilon_5+\varepsilon_2-\varepsilon_3} \bigg],\nonumber
\end{eqnarray}
where $n_i\equiv n^0(\varepsilon_i)$ are the Bose distribution
functions with the corresponding energies and
$n_{i-j}\equiv n^0(\varepsilon_i-\varepsilon_j)$.

For the uniform, DC thermal conductivity (\ref{K_kappa0}) we need to take the imaginary part and the $\omega\rightarrow 0$ limit of (\ref{IAB_3a}):
${\rm Im}\,
\overline{\Pi}_{A}(i\omega_n\rightarrow\omega+i0)$ at $\omega\rightarrow 0$, which splits naturally into four terms:
\begin{eqnarray}
\label{IAB_4}
&&{\rm Im}\,\overline{\Pi}_{A}(\omega)=\sum_{m=1}^4
{\rm Im}\,\overline{\Pi}_{A}^{(m)}(\omega)\\
&&=
{\rm Im}\,({\rm I}) \, {\rm Re}\,({\rm II})\, {\rm Re}\,({\rm [III]})
+{\rm Re}\,({\rm I}) \, {\rm Im}\,({\rm II})\,
{\rm Re}\,({\rm [III]})\nonumber\\
&&
+{\rm Re}\,({\rm I}) \, {\rm Re}\,({\rm II})\, {\rm Im}\,({\rm [III]})
-{\rm Im}\,({\rm I}) \, {\rm Im}\,({\rm II})\, {\rm Im}\,({\rm [III]})\, ,
\nonumber
\end{eqnarray}
where ${\rm I}$, ${\rm II}$,  and ${\rm [III]}$ are
the first, second, and third factors of the product in (\ref{IAB_3a}),
respectively, and ${\rm [III]}$ includes all the terms inside the square bracket. In what follows, we refer to
the four contributions to the conductivity
coming from the four terms in (\ref{IAB_4}) as to ${\rm Im}\Pi_A^{(m)}$,
$m=1\dots 4$.

\subsubsection{Boltzmann terms}
\label{BLT}

Here we explicitly evaluate the leading-order contributions to the thermal drag conductivity and show that the Kubo approach yields the same answer as
the one obtained using Boltzmann equation. The
discussion of the other, subleading non-Boltzmann contributions is deferred to Appendix~\ref{sec:NBC}.

We would like to assert that within the spectral representation calculation, the leading contributions are given only by
${\rm Im}\,\overline{\Pi}_{A}^{(4)}$ in (\ref{IAB_4}).
The rest of the terms, ${\rm Im}\Pi_A^{(1)}$, ${\rm Im}\Pi_A^{(2)}$, and  ${\rm Im}\Pi_A^{(3)}$, yield results that are subleading in the sense of containing higher power of $\tau$'s, which is equivalent to having higher-order terms in $V^{\rm b-ph}$ and other couplings.

Consider ${\rm Im}\,\overline{\Pi}_{A}^{(4)}$ term
where all three factors in (\ref{IAB_3a}) contribute their imaginary parts,
\begin{eqnarray}
\label{B4_0}
&&{\rm Im}\,\overline{\Pi}_{A}^{(4)}(\omega)=\pi^3 \delta(\omega+\varepsilon_1-\varepsilon_2) \delta(\omega+\varepsilon_5-\varepsilon_4)\nonumber\\
&&\phantom{{\rm Im}}
\big[(n_3-n_1)(n_{3-1}-n_5)
\delta(\omega+\varepsilon_5+\varepsilon_1-\varepsilon_3)
\\
&&\phantom{{\rm Im}}
-(n_3-n_2)(n_{3-2}-n_4)
\, \delta(\varepsilon_4+\varepsilon_2-\varepsilon_3-\omega)
\big].\nonumber
\end{eqnarray}
In the low frequency limit Eq.~(\ref{B4_0}) reduces to:
\begin{eqnarray}
\label{B4_1}
&&
\frac{{\rm Im}\,\overline{\Pi}_{A}^{(4)}(\omega)}{\omega}\Big|_{\omega=0}
=2\pi^3 \delta(\varepsilon_1-\varepsilon_2) \delta(\varepsilon_5-\varepsilon_4)\nonumber\\
&&\phantom{\frac{{\rm Im}\,\overline{\Pi}_{A}^{(4)}(\omega)}{\omega}} \times(n_3-n_1)\frac{\partial n_5}{\partial \varepsilon}\delta(\varepsilon_5+\varepsilon_1-\varepsilon_3).
\end{eqnarray}
Substituting this into (\ref{IAB_1}) and performing
integrations with delta-functions, we
 obtain the leading contribution to
${\rm Im}\,\Pi_{A}({\bf k},{\bf p},\omega)$,
\begin{eqnarray}
\label{B4_2}
&&\frac{{\rm Im}\,
\Pi_{A}^{(4)}
({\bf k},{\bf p},\omega)}{\omega}\bigg|_{\omega=0}=
\frac14
\int_{\varepsilon_1}\Big(A_{\bf k}(\varepsilon_1)\Big)^2
\int_{\varepsilon_5}\Big({\bar A}_{\bf p}(\varepsilon_5)\Big)^2
\nonumber\\
&&\phantom{{\rm Im}\, \frac{I_{A}^{(\#4)}(\omega)}{\omega}\bigg|_{\omega=0}}
A_{\bf k-p}(\varepsilon_1+\varepsilon_5)
\ [n_{1+5}-n_1]\,
\frac{\partial n_5}{\partial\varepsilon}\,
.
\end{eqnarray}
We assume that bosons and phonons are
well-defined quasiparticles with frequency-independent imaginary
parts of their self-energies, $r_{\bf k}$ and $s_{\bf p}$,
such that $r_{\bf k} (s_{\bf p})\ll \varepsilon_{\bf k}
(\omega_{\bf p})$. They are also related to the relaxation times used in
Sec.~\ref{Boltzmann} as $r_{\bf k}^{-1}=2\tau^{\rm b}_{\bf k}$ and $s_{\bf
p}^{-1}=2\tau^{\rm ph}_{\bf p}$, see Ref.~\onlinecite{mahan1990}.  Thus, the
spectral functions of bosons and phonons can be approximated as Lorentzians:
\begin{eqnarray}
\label{A1}
&&A_{\bf k}(\varepsilon_1)=
\frac{2r_{\bf k}}
{(\varepsilon_1-\varepsilon_{\bf k})^2+r_{\bf k}^{2}}\, ,\\
\label{A2}
&&{\bar A}_{\bf p}(\varepsilon_5)=
\frac{2s_{\bf p}}
{(\varepsilon_5-\omega_{\bf p})^2+s_{\bf p}^{2}}\,.
\end{eqnarray}
Since the spectral functions (\ref{A1}) and (\ref{A2})
are  strongly peaked at $\varepsilon_{\bf k}$
and $\omega_{\bf p}$,   the main contributions in the
integrals in (\ref{B4_2}) are obtained at
$\varepsilon_1\approx \varepsilon_{\bf k}$ and $\varepsilon_5\approx
\omega_{\bf p}$.
Identifying distribution functions with that of phonons and bosons via
$n(\omega_{\bf p})\equiv n^0_{\bf p}$,
$n(\varepsilon_{\bf k})=f^0_{\bf k}$, and
$n(\varepsilon_{\bf k-p})=f^0_{\bf k-p}$,
 finally yields,
\begin{eqnarray}
\label{B4_4}
&&\frac{{\rm Im}\,
\Pi_{A}({\bf k},{\bf p},\omega)}{\omega}\bigg|_{\omega=0}\approx
\, 2\pi\,
\tau^{\rm b}_{\bf k}\, \tau^{\rm ph}_{\bf p}\,
\frac{\partial
n^0_{\bf p}}{\partial \omega}\big[f^0_{\bf k-p} -f^0_{\bf k}\big]
\nonumber\\
&&\phantom{{\rm Im}\, \frac{I_{A}^{(\#4)}(\omega)}{\omega}\bigg|_{\omega=0}=
\, 2\pi}
\times
A_{\bf k-p}(\varepsilon_{\bf k}+\omega_{\bf p}) \\
&& \approx
 2\pi\, \tau^{\rm b}_{\bf k}\, \tau^{\rm ph}_{\bf p}\,
\frac{\partial n^0_{\bf p}}{\partial \omega}\big[f^0_{\bf k-p} -f^0_{\bf k}\big]
\cdot \delta(\varepsilon_{\bf k}+\omega_{\bf p}-\varepsilon_{\bf k-p}),\nonumber
\end{eqnarray}
where in the last line we have approximated the Lorenzian with the
delta-function and have neglected contributions from $\Pi_{A}^{(1-3)}$ terms.  As we discuss in Appendix~\ref{sec:NBC}, both approximation are of the same order and correspond to neglecting terms that are subleading to (\ref{B4_4}).

Repeating the same consideration for  the diagram
$B$ in Fig.~\ref{diagrams} gives:
\begin{eqnarray}
\label{A4_4}
&&\frac{{\rm Im}\,
\Pi_{B}({\bf k},{\bf p},\omega)}{\omega}\bigg|_{\omega=0}\approx
\, 2\pi\,
\tau^{\rm b}_{\bf k}\, \tau^{\rm ph}_{\bf p}\,
\frac{\partial
n^0_{\bf p}}{\partial \omega}\big[f^0_{\bf k-p} -f^0_{\bf k}\big]
\nonumber\\
&&\phantom{{\rm Im}\, \frac{I_{A}^{(\#4)}(\omega)}{\omega}\bigg|_{\omega=0}=
\, 2\pi}
\times
\delta(\varepsilon_{\bf k}-\omega_{\bf p}-\varepsilon_{\bf k-p})
\,
.
\end{eqnarray}
Using the identities  for the distribution
functions in (\ref{convert1}), (\ref{convert2}) and
substituting (\ref{B4_4}) and (\ref{A4_4}) into (\ref{K_kappa0})
gives the Kubo answer for the drag component of the thermal conductivity
\begin{eqnarray}
\label{K_kappa_4}
&&\kappa_{\rm drag}=-\frac{4\pi}{T}\sum_{\bf k,p}
\left(v^x_{\bf k}\,\varepsilon_{\bf k}\,\tau^{\rm b} _{\bf k}\right)\,
\left(u^x_{\bf p}\,\omega_{\bf p}\,\tau^{\rm ph}_{\bf p}\right)\,
\big|V^{\rm b-ph}_{{\bf p};{\bf k},{\bf k}-{\bf p}}\big|^2
\nonumber\\
&&\phantom{\kappa_{\rm drag}}
\times\bigg[\frac{\partial f^0_{\bf k}}{\partial \varepsilon}\,
\big[f^0_{\bf k-p} +n^0_{\bf p}+1\big]
\delta\big(\varepsilon_{\bf k}-
\varepsilon_{\bf k-p}-\omega_{\bf p}\big)\ \ \ \ \ \\
&&\phantom{\kappa_{\rm drag}=}
-\frac{\partial f^0_{\bf k}}{\partial \varepsilon}\,
\big[f^0_{\bf k-p} -n^0_{\bf p}\big]
\delta\big(\varepsilon_{\bf k}-
\varepsilon_{\bf k-p}+\omega_{\bf p}\big)
\bigg]\, .\nonumber
\end{eqnarray}
One can see that this is identical to the Boltzmann answer in (\ref{kappatotal}).
We discuss in Appendix~\ref{sec:NBC} that the
contributions from  the remaining terms ${\rm Im}\Pi_A^{(m)}$, $m=1\dots 3$
and corrections to (\ref{K_kappa_4}) due to the broadening
in the spectral functions are of the  order $O(\tau^3)$ and, therefore, can
 be neglected.

One can check the consistency of the drag conductivity expression in (\ref{kappatotal}) and (\ref{K_kappa_4}) with the diagonal terms in conductivity by assuming that the leading source of the relaxations defining both the spin and phonon transport relaxation times is the spin-phonon coupling in (\ref{Hint}). Then the diagonal
and the drag conductivities are all of the same order in the spin-phonon coupling:  $\kappa^{\rm ph}\sim \kappa^{\rm b}\sim\kappa_{\rm drag}\propto 1/|{\widetilde V}^{\rm b-ph}|^2$. This also demonstrates that in the idealized case of free spin excitations coupled to dissipationless phonons via a weak coupling all conductivities should be of the same order in
that coupling.

We would like to emphasize again, that the identity between
Eq.~(\ref{K_kappa_4}) and Eq.~(\ref{kappatotal}) is rather remarkable as they
are derived starting from completely different physical formulations.

\section{fermions and other}
\label{other}

Here we generalize the analysis of this work onto two additional cases. First,
we assume the same form of the coupling to phonons (\ref{Hint}), but
consider fermions instead of bosons.  This scenario is not only
applicable to cases where the spin algebra has been mapped onto
fermions, but is also relevant to the thermal conductivity in metals and
semiconductors. The second generalization
extends our drag consideration on the case of anomalous bosonic terms in the boson-phonon interaction.
Such terms readily exist in the interaction of phonons with magnons in the
ordered antiferromagnets, as was discussed previously.\cite{Dixon} They also exist for triplet excitations in gapped, dimerized, and other
phases.  The following derivations are based on the Boltzmann formalism
only.

\subsection{fermions}

Coupling of fermionic excitations with phonons is, generally, of the same form
as given in Eq.~(\ref{Hint}). This is obviously the case for the coupling of
electrons with phonons, and is also true for the $XXZ$ spin chains when spins
are represented by the Jordan-Wigner fermions.
The general expression for the drag current will be still given by
Eq.~(\ref{j3}), where $\varepsilon_{\bf k}$ is replaced by
$\widetilde{\varepsilon}_{\bf k} =\varepsilon_{\bf k}-\mu$, the fermion energy
relative to the chemical potential, and the ``drag rates'' $1/\tau^{\rm
ph\rightarrow f}$ and $1/\tau^{\rm f\rightarrow ph}$ are determined by the
corresponding collision integral involving fermions and bosons, with $f$ now
representing the fermion occupation number and $f^0_{\bf k}
=(\exp[(\varepsilon_{\bf k}-\mu)/T]+1)^{-1}$ being the equilibrium
Fermi-distribution function.  One obvious difference for the probabilities is
that a fermion with the momentum $\bf{k}$ is created with the probability given
by $(1 - f_{\bf k})$.  The derivation for the drag conductivity follows exactly
the same steps as those for the boson-phonon case considered in Section
\ref{Boltzmann}.  Useful identities for certain combinations of $f^0$ and
$n^0$, analogous to the ones in (\ref{convert1}) and (\ref{convert2}),
 are listed
in Appendix~\ref{app_A}.2.  Taking into consideration the above differences we
obtain the total drag conductivity
\begin{eqnarray}
\label{kappaphf}
&&\kappa_{\rm drag}=-\frac{4\pi}{T}\sum_{\bf k,p}
\left(v^x_{\bf k}\,\widetilde{\varepsilon}_{\bf k}\,\tau^{\rm f} _{\bf k}\right)\,
\left(u^x_{\bf p}\,\omega_{\bf p}\,\tau^{\rm ph}_{\bf p}\right)\,
\big|V^{\rm f-ph}_{{\bf p};{\bf k},{\bf k}-{\bf p}}\big|^2
\nonumber\\
&&\phantom{\kappa=}
\times\Bigg[\frac{\partial f^0_{\bf k}}{\partial \varepsilon}\,
\big[1-f^0_{\bf k-p} +n^0_{\bf p}\big]\delta\big(\varepsilon_{\bf k}-
\varepsilon_{\bf k-p}-\omega_{\bf p}\big)
\ \ \ \ \ \\
&&\phantom{\kappa=\times\Bigg[}
+\frac{\partial f^0_{\bf k}}{\partial \varepsilon}\,
\big[f^0_{\bf k-p} +n^0_{\bf p}\big]\delta\big(\varepsilon_{\bf k}-
\varepsilon_{\bf k-p}+\omega_{\bf p}\big)\Bigg].\ \ \ \ \ \nonumber
\end{eqnarray}
To summarize, the phonon drag conductivity for the fermionic
case (\ref{kappaphf})
takes the same form as for the bosonic case (\ref{kappatotal}) with two
modifications: (i) $f^0_{\bf k-p}\rightarrow -f^0_{\bf k-p}$,
(ii) $\varepsilon_{\bf k}\rightarrow
\widetilde{\varepsilon}_{\bf k}=\varepsilon_{\bf k}-\mu$.
Note that the second change should also be made in the case of bosons
if the chemical potential for them is not zero.

We note, that the fermionic case of the drag discussed here is different from the one traditionally considered in the thermoelectric phenomena. 
As is mentioned in Sec.~\ref{intro}, for the electron-phonon problem in metals, the thermal-only drag effect is usually neglected because of the dominance of the electronic thermal conductivity over the phonon one.\cite{Ziman}
This is {\em not} the case in many low-dimensional quantum magnets.
\cite{Hess2001,Sologubenko2001,Sologubenko2003,Ribeiro2005,Hess2007,Hess2010}

 Regarding potentially different outcomes of the drag effect for the fermionic systems (\ref{kappaphf}) compared to the bosonic ones
(\ref{kappatotal}) [(\ref{K_kappa_4})], we remark that the major difference
may arise due to the presence of the Fermi surface in the former cases. 
It is known, that the ``normal''
and the Umklapp scatterings contribute with opposite sign to the drag
conductivity, which is discussed as one of the reasons for the suppression of
the Gurevich effect in metals.\cite{Ziman,Ziman_eph,thermpwr} Such an effect of Umklapp can
be expected to be small at low temperature for the bosonic case because all of
the heat carriers are at small momenta. For the fermionic case, on the other
hand, the effect of the Umklapp should be present, similarly to the
electron-phonon case.  However, since the fermionic representation of spins is
restricted to 1D, significant differences from the traditional 3D
electron-phonon consideration may also occur. Any quantitative statement on whether
the drag will be more substantial for magnetic excitations obeying bosonic or
fermionic statistics will depend on specific model calculations, which are not
the focus of this work.

\subsection{anomalous bosonic terms}

Next we consider drag contributions in the
phonon-boson case due to anomalous terms of the kind
\begin{eqnarray}
\label{Hint1}
{\cal H}= \sum_{{\bf k},{\bf p}}
{\widetilde V}^{\rm b-ph}_{{\bf p};{\bf k},{\bf k}-{\bf p}}
\ \left(b^\dag_{\bf -k+p}b^{\dag}_{\bf k}a_{\bf p}
+{\rm H.c.}\right).
\end{eqnarray}
These describe processes involving creation of two bosons from a phonon and
generation of a phonon due to the annihilation of two bosons.

With the details of the algebra provided in Appendix~\ref{app_A}.3, here we
simply state that the approach described in Sec.~\ref{Boltzmann} yields the
following result
\begin{eqnarray}
&&\kappa_{\rm drag}=-\frac{4\pi}{T}\sum_{\bf k,p}
\left(v^x_{\bf k}\,\varepsilon_{\bf k}\,\tau^{\rm f} _{\bf k}\right)\,
\left(u^x_{\bf p}\,\omega_{\bf p}\,\tau^{\rm ph}_{\bf p}\right)\,
\big|{\widetilde V}^{\rm b-ph}_{{\bf p};{\bf k},{\bf k}-{\bf p}}\big|^2
\nonumber \\
&&\phantom{\kappa=}
\times\frac{\partial n^0_{\bf k}}{\partial \omega}\,
\big[1+ f^0_{\bf k}  + f^0_{\bf k-p}\big]\delta\big(\varepsilon_{\bf k}+
\varepsilon_{\bf k-p}-\omega_{\bf p}\big)
\ . \hphantom{aaa}\label{kappaphf4}
\end{eqnarray}
In the case when both the ``normal'' (\ref{Hint}) and ``anomalous'' (\ref{Hint1}) boson-phonon couplings are present, the leading contribution
to the drag thermal conductivity is the sum of the results in (\ref{kappatotal}) and (\ref{kappaphf4}).

\section{Qualitative estimates}
\label{sec:qualitative}

In this section we provide a qualitative
discussion of various asymptotic results that can be readily inferred from Eq.~(\ref{kappatotal}) for several representative spin-phonon systems
with the goal of estimating
when drag effects can be significant and when they are not.
The temperature dependence of the drag thermal conductivity is
determined by two factors:  scattering lifetimes and the
occupation numbers of the excitations.

\subsection{Boundary-limited regime}

First, we would like to consider gapless spin excitations with linear dispersion $\varepsilon_{\bf k}\approx v|{\bf k}|$, coupled to acoustic 3D phonons, the situation relevant to a wide variety of antiferromagnets.
For the low impurity concentration and at low temperatures both phonon and boson mean-free paths can be expected to be boundary-limited, the case well documented for Nd$_2$CuO$_4$.\cite{Taillefer_08}
However, the heat carrying excitations will be few in number and the drag conductivity has to go to zero at low temperatures.
A straightforward algebra in (\ref{kappatotal}) yields a power-law:
$\kappa_{\rm drag}\propto T^{\gamma}$, with $\gamma=2+D_s+m$, where $D_s$
is the dimensionality of the spin system and $m$ depends on the
long-wavelength ${\bf k}$- and ${\bf p}$-dependence of the spin-phonon coupling
${\widetilde V}^{\rm b-ph}_{{\bf p};{\bf k},{\bf k}-{\bf p}}$. In the case of $D_s=3$ (e.g., 3D magnons) and assuming that the coupling follows the standard form ${\widetilde V}^{\rm b-ph}_{{\bf p};{\bf k},{\bf k'}}\propto
\sqrt{pkk'}$, which corresponds to $m=3$, altogether
gives  $\kappa^{3D}_{\rm drag}\propto T^8$. This should be compared with the diagonal thermal conductivities in this regime
$\kappa^{\rm ph}\sim \kappa^{\rm b}\propto T^3$. Thus, the drag effect is, generally, subleading in the considered regime.

\subsection{Gapped spin system}

In another specific
example let us consider a gapped spin system  at low enough temperatures so that the occupation number of spin excitations is exponentially small: $f_{\bf k}\propto e^{-\Delta/T}$, where $\Delta$ is the gap in the
spectrum. In the case when the relaxation within the spin system is only due to
a weak coupling to phonons whose relaxation rate is dominated by the Umklapp processes, i.e. $\tau^{\rm ph}_{\bf p}\propto
e^{\widetilde\Theta_D/T}$, where $\widetilde\Theta_D$ is a
fraction of the Debye energy,\cite{Berman}
 our Eq.~(\ref{kappatotal}) naturally leads to
$\kappa_{\rm drag}\propto e^{(\widetilde\Theta_D -\Delta)/T}$.
This result was obtained in Ref.~\onlinecite{Boulat2007} using the
memory-matrix approach.

\subsection{High-temperature regime and disorder effects}

Third, we consider the high-temperature limit, for either gapped or gapless spin
system, when temperature is higher than both the Debye energy and the
spin-excitation energy scale, $T\gg \Theta_D, J$. Formally this case may raise
questions regarding the transition into the disordered state, however is fully
analogous to the textbook consideration of the lattice thermal conductivity at
$T\gg \Theta_D$.\cite{Ziman} In fact. in this region quasiparticles can be
considered as strongly damped. The rates of the Umklapp scattering for spin
excitations and phonons are high and are proportional to the occupation numbers
of a ``typical'' boson or phonon, thus leading to $\tau^{\rm ph}\sim\tau^{\rm
b}\propto 1/T$. The rest of the estimate in Eq.~(\ref{kappatotal}) is again
straightforward, giving $\kappa_{\rm drag}\propto 1/T$. This should be compared
to the diagonal conductivities in this regime, which show the same asymptotic
behavior:\cite{footnote1} $\kappa^{\rm ph}\sim \kappa^{\rm b}\propto 1/T$.  This
consideration, combined with the low-temperature one, implies that the drag
conductivity should go through a maximum at intermediate temperatures, similar
to the diagonal conductivities.

When the energy scales of the phonon and spin system are well separated, as
in the cuprate-based materials where $J\gg\Theta_D$, another asymptotic regime is possible, $\Theta_D\ll T\ll J$. Intuitively, the drag can be expected to diminish together with the phonon conductivity
($\kappa^{\rm ph}\propto 1/T$)
because phonons are sufficiently equilibrated by the phonon-phonon scattering.
However, the $T$-dependence of the drag also depends on the specifics of the relaxation within the spin system. Thus, no definite conclusion on the prevalent behavior of the drag conductivity in this regime can be drawn without identifying such a relaxation.

The disorder dependence of the drag can also be considered using similar qualitative reasoning. If the disorder affects both types of excitations on equal footing, so that $\kappa^{\rm ph}\sim \kappa^{\rm b}\propto 1/n_{\rm imp}$, where
$n_{\rm imp}$ is the impurity concentration, then the drag conductivity diminishes as $\kappa_{\rm drag}\propto 1/(n_{\rm imp})^2$.
If the disorder can be introduced
selectively in one of the sub-systems without significantly affecting the other, as in the case of  lattice disorder in the ladder cuprate system Ca$_9$La$_5$Cu$_{24}$O$_{41}$,\cite{Hess2001} the drag conductivity will be reduced together with the diagonal conductivity of the most affected species of excitations.

Thus, intuitive conditions for maximizing the effect of drag
are the simultaneous presence of significant population of spin excitations
and phonons with long scattering times. Since such conditions also imply
large diagonal contributions of spins and phonons to the heat current, they
are typically satisfied for temperatures that are low enough in comparison with either $J$ or $\Theta_D$ but are above the boundary-limited regime.
Note that the optimal regime for the phonon drag in thermoelectric phenomenon is
often quoted as $T\sim\Theta_D/5$.\cite{Ziman_eph}
Such a regime can be of relevance to the recently reported record-breaking thermal conductivity by spin excitation in a high-purity
1D spin-chain material SrCuO$_2$, Ref.~\onlinecite{Hess2010}, where a nearly ballistic propagation of spin excitations was reported.

The issue of the separation of the drag component of the thermal
conductivity from the ``diagonal'' one may require a series of doping
experiments in which disorder is introduced deliberately to suppress the
conductivity of one of the species and thus diminishing the drag as
well.\cite{Hess2010}

\subsection{qualitative estimate of the drag}

Lastly, we would like to come back to the problem of the
gapless spin excitations with linear dispersion coupled to phonons,
the problem motivated by the 1D spin-chain and 2D layered cuprates where the spin excitations are fast and the phonons are slow, $J\gg\Theta_D$.
Analogous to similar estimates of the thermoelectric power,\cite{Ziman}
and without reference to a specific model, the following consideration is not
intended to be entirely rigorous, but rather is aimed at deriving an upper-limit
estimate of the thermal Gurevich effect.

Let us assume that the boson relaxation is due to impurities or some other extrinsic or intrinsic mechanism while phonons are dissipationless,
a consideration similar to the electron-phonon drag problem.\cite{Ziman}
Such a scenario is also
potentially relevant to the 1D spin-chain materials in low-$T$ regime.
Then, the spin-phonon coupling will provide both the dissipation for phonons and the drag between phonons and spin excitations. In the drag conductivity, the phonon relaxation time ($\tau^{\rm ph}\propto 1/|{\widetilde V}^{\rm b-ph}|^2$)
enters together with the ``drag rates''
($1/\tau^{\rm ph\leftrightarrow b}\propto |{\widetilde V}^{\rm b-ph}|^2$). As shown in Appendix \ref{app_A}.4, one can demonstrate that for quasiparticles with linear dispersions and for ${\bf k}$-independent boson relaxation time $\tau^{\rm b}_{\bf k}=\tau^{\rm b}$ the following simplification for the drag conductivity is possible for an {\it arbitrary} form of the coupling
${\widetilde V}^{\rm b-ph}_{{\bf p};{\bf k},{\bf k}-{\bf p}}$:
\begin{eqnarray}
\label{simplify}
\kappa_{\rm drag} = -\frac{u^2 v^2\tau^{\rm b}}{T}
\sum_{\bf p} \frac{\partial n^0_{\bf p}}{\partial \omega}
\left({\bf p}^x\right)^2 = \frac{1}{3}v^2 \tau^{\rm b} C_{\rm ph}\, ,
\end{eqnarray}
where $u$ and $v$ are the phonon and boson velocities, and $C_{\rm ph}$ is the phonon specific heat. Since the diagonal conductivity of bosons in this case is
$\kappa^{\rm b}= \frac{1}{D_s}v^2 \tau^{\rm b} C_{\rm b}$, where $D_s$ is the dimensionality of the spin system, the ratio of the drag conductivity to the boson one is independent of the scatterings and is defined by the boson and phonon specific heats
\begin{eqnarray}
\label{Ratio}
\frac{\kappa_{\rm drag}}{\kappa_{\rm b}} =\frac{D_s}{3}\, \cdot\,
\frac{C_{\rm ph}}{C_{\rm b}}\, .
\end{eqnarray}
Since the population of phonons at a given temperature can be much larger than  that of bosons, the drag conductivity can significantly exceed the one by spin excitations.
Similar argument is at the core of the original proposal by Gurevich for the large thermoelectric effect in metals,\cite{Ziman,Gurevich46} where the relation
$\kappa_{\rm drag}/\kappa_{\rm e} = C_{\rm ph}/C_{\rm e}$ also implies the same drift velocities of phonons and electrons.

While the parallel and the similarity between the electron-phonon drag and the thermal-only drag considered in the last example 
are clear, they are not complete.
The difference is in the presence of another diagonal conductivity term in our consideration, $\kappa^{\rm ph}\propto 1/|{\widetilde V}^{\rm b-ph}|^2$, which, in the limit of the small spin-phonon coupling will dominate both
$\kappa_{\rm drag}$ and $\kappa_{\rm b}$.  Therefore, in general, the relation (\ref{Ratio})  {\it does not} imply the equivalence of the drift velocities of phonons and spin excitations.

\section{Conclusion}
\label{sec:Conclusion}

In this work we have considered a two-component system
of phonons and spin excitations and have
obtained general expression for the off-diagonal contribution
to its thermal conductivity in the lowest order of
the spin-phonon coupling.
The off-diagonal contribution to the thermal
current, referred to as  thermal drag,
is an enhancement of the heat flux of one of the
species due to the flow of another and vice-versa.  We have
employed two distinct approaches, the Boltzmann formalism
and the Kubo approach, to derive the spin-phonon drag thermal conductivity
and have established that both approaches yield identical
results, Eqs.~(\ref{kappatotal}) and (\ref{K_kappa_4}).
In addition, we have considered contributions to drag from anomalous
terms, which generally arise from the spin-phonon coupling, e.g. in the symmetry broken phases as well as in the gapped systems characterized by triplet-like excitations.
 While we mainly focus on the drag between phonons and bosonic
spin excitations, we have also discussed the case where the spin excitation's
statistics is fermionic.

To conclude, we have obtained an explicit
expression for the drag conductivity in the two-component system of phonons and spin excitations  under general assumptions on the nature of interaction between them. This should allow for the practical calculations of the drag effects in a number of materials.

\begin{acknowledgments}

This work has been initiated at the Kavli Institute of Theoretical Physics
and part of this work has been done at the Max Planck Institute for the Physics
of Complex Systems, and the Aspen Institute of Physics, which we would like
to thank for hospitality.  This work was supported by DOE under grant
DE-FG02-04ER46174 (A. L. C.) and by the DFG through Grant No. BR 1084/6-1 of
FOR912 (W. B.). The research at KITP was supported by the NSF under Grant
No. PHY05-51164.

\end{acknowledgments}

\appendix
\section{Details of the Boltzmann approach}\label{sec:DetailsBZA}
\label{app_A}

In this Appendix we provide some further details of the Boltzmann approach to
the drag discussed in Secs.~\ref{Boltzmann} and \ref{other}.

\subsection{Derivation of the drag rate of phonons on bosons}
Here we derive the ``drag rate'' of phonons on bosons
$1/\tau^{\rm ph\rightarrow b}_{{\bf p}, {\bf k}}$ in Eq.~(\ref{BE0a}).

\begin{figure}[t]
\centering
\includegraphics[width=1.0\columnwidth]{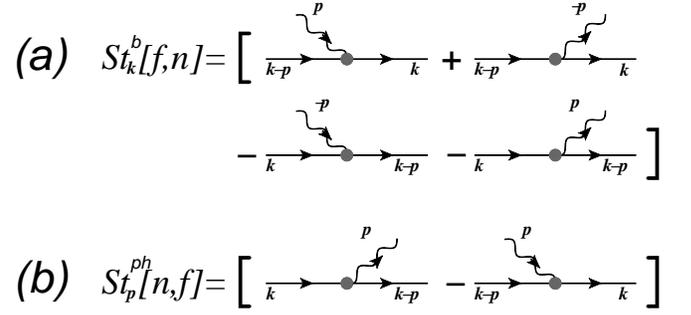}
\caption{Graphical representation of the collision integral terms.}
\label{St1_diag}
\end{figure}

While not necessary, it is nevertheless convenient to depict processes
contributing to the collision integral as ``probability diagrams'', see
Fig.~\ref{St1_diag}.  The collision integral for bosons scattered off
phonons via interaction (\ref{Hint}) contains four terms, see
Fig.~\ref{St1_diag}(a): the first two increase the number of bosons with
momentum ${\bf k}$ the other two scatter ${\bf k}$-bosons into a different
state. They can be grouped together by energy-conservation to yield,
\begin{eqnarray}
\label{Stf}
&&St^{\rm b}_{\bf k}[f,n]=2\pi\sum_{\bf p}
\big|V^{\rm b-ph}_{{\bf p};{\bf k},{\bf k}-{\bf p}}\big|^2\nonumber\\
&&\phantom{St^{\rm b,ph}_{\bf k}}
\Big(\big[f_{\bf k-p} n_{\bf p}(f_{\bf k}+1)
-f_{\bf k} (n_{\bf p}+1)(f_{\bf k-p}+1)\big]\\
&&\phantom{\Big(\big[f_{\bf k-p} n_{\bf p}(f_{\bf k}+1)
-f_{\bf k}(n_{\bf p}+}
\times\delta\big(\varepsilon_{\bf k}-
\varepsilon_{\bf k-p}-\omega_{\bf p}\big)\nonumber\\
&&\phantom{St^{\rm b,ph}_{\bf k}}
+\big[f_{\bf k-p} (n_{\bf -p}+1)(f_{\bf k}+1)
-f_{\bf k} n_{\bf -p}(f_{\bf k-p}+1)\big]\nonumber\\
&&\phantom{\Big(\big[f_{\bf k-p} n_{\bf p}(f_{\bf k}+1)
-f_{\bf k}(n_{\bf p}+}
\times\delta\big(\varepsilon_{\bf k}-
\varepsilon_{\bf k-p}+\omega_{\bf p}\big)\Big).\nonumber
\end{eqnarray}
using $V_{-\mathbf{p};\mathbf{k}-\mathbf{p},\mathbf{k}}^{\rm b-ph}
=(V_{\mathbf{p};\mathbf{k},\mathbf{k}-\mathbf{p}}^{\rm b-ph})^{\star}$ discussed
after (\ref{Hint}), writing $f=f^0+\delta f$ and $n=n^0+\delta n$ and neglecting
terms of order $\delta f\delta n$ and $\delta f\delta f$ will yield
the terms proportional to $\delta f$ and $\delta n$ shown in
Eq.~(\ref{BE0a}).  Performing this procedure and using
$\delta n_{\bf
-p}=-\delta n_{\bf p}$ yields the drag rate of phonons on bosons:
\begin{eqnarray}
\label{taunf}
&&St^{\rm b}_{\bf k}[f^0,n^0+\delta n]\approx
-\sum_{\bf p}\frac{\delta n_{\bf p}}{\tau^{\rm ph\rightarrow b}_{\bf p,k}}
,\nonumber\\
&&\frac{1}{\tau^{\rm ph-b}_{\bf p,k}}=-2\pi
\big|V^{\rm b-ph}_{{\bf p};{\bf k},{\bf k}-{\bf p}}\big|^2
\\
&&\phantom{\frac{1}{\tau^{\rm ph-b}_{\bf p,k}}=}
\Big(\big[f^0_{\bf k-p} -f^0_{\bf k}\big]
\delta\big(\varepsilon_{\bf k}-
\varepsilon_{\bf k-p}-\omega_{\bf p}\big)\nonumber\\
&&\phantom{\frac{1}{\tau^{\rm ph-b}_{\bf p,k}}=}
-\big[f^0_{\bf k-p} -f^0_{\bf k}\big]
\delta\big(\varepsilon_{\bf k}-
\varepsilon_{\bf k-p}+\omega_{\bf p}\big)\Big).\nonumber
\end{eqnarray}
Substituting this into the thermal conductivity in Eq.~(\ref{kappa}) and using
relations (\ref{convert1}) and (\ref{convert2}) yields the thermal conductivity
in (\ref{kappa2equal}) and (\ref{kappatotal}).

The phonon collision integral discussed in Sec.~\ref{Boltzmann}.B is shown
in Fig.~\ref{St1_diag}(b).

\subsection{Useful identities for the fermionic case}

Identities similar to Eqs.~(\ref{convert1}) and (\ref{convert2}) that are useful
for simplifying  expressions for the thermal conductivity and for
relating its components to each other can also be obtained for the
fermion-phonon system. They are
\begin{eqnarray}
\label{convert1f}
\frac{\partial n^0_{\bf p}}{\partial \omega}
\big[f^0_{\bf k-p} -f^0_{\bf k}\big]\equiv
\frac{\partial f^0_{\bf k}}{\partial \varepsilon}
\big[n^0_{\bf p}-f^0_{\bf k-p}+1\big]\Bigg|_{\omega_{\bf p}=\varepsilon_{\bf k}-\varepsilon_{\bf k-p}}\\
\label{convert2f}
\frac{\partial n^0_{\bf p}}{\partial \omega}
\big[f^0_{\bf k} -f^0_{\bf k-p}\big]\equiv
\frac{\partial f^0_{\bf k}}{\partial \varepsilon}
\big[f^0_{\bf k-p} +n^0_{\bf p}\big]\Bigg|_{\omega_{\bf p}=
\varepsilon_{\bf k-p}- \varepsilon_{\bf k}}.\ \ \ \
\end{eqnarray}
These identities help to see that both contribution to the drag
are equivalent.

\subsection{Derivation of the drag rates for the anomalous boson-phonon coupling}

\begin{figure}[t]
\centering
\includegraphics[width=0.99\columnwidth]{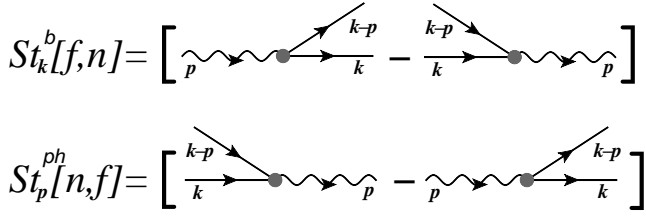}
\caption{Graphical representation of the collision integrals for the ``anomalous'' terms, Eq.~(\ref{Hint1}).}
\label{St1a_diag}
\end{figure}

The derivation of the drag in the case of the anomalous boson-phonon
coupling is similar to the procedure detailed in Sec.~\ref{Boltzmann} and Appendix \ref{app_A}.1.  For the coupling in (\ref{Hint1}) the
scatterings describe the processes involving creation of two bosons from a
phonon and generation of a phonon due to annihilation of two bosons.
In that case the boson and the phonon collision integrals are described by two
similar ``probability diagrams'', see Fig.~\ref{St1a_diag}.

The expression for the boson collision integral has the following form,
\begin{eqnarray}
\label{Stfph1}
&&St^{\rm b}_{\bf k}[f,n]=2\pi\sum_{\bf p}
\big|{\widetilde V}^{\rm b-ph}_{{\bf p};{\bf k},{\bf k}-{\bf p}}\big|^2
\cdot \delta\big(\varepsilon_{\bf k}-
\varepsilon_{\bf k-p}-\omega_{\bf p}\big)\ \ \ \ \
\nonumber\\
&&\phantom{St^{\rm b}}
\times\big[(1+f_{\bf k-p}) (1+f_{\bf k}) n_{\bf p}
-f_{\bf k-p} f_{\bf k} (1 + n_{\bf p})\big].\ \
\end{eqnarray}
Thus, the drag rate of phonons on bosons is:
\begin{eqnarray}
\label{taunf_a}
&&St^{\rm b}_{\bf k}[f^0,n^0+\delta n]\approx
-\sum_{\bf p}\frac{\delta n_{\bf p}}{\tau^{\rm ph\rightarrow b}_{\bf p,k}}
,\nonumber\\
&&\frac{1}{\tau^{\rm ph\rightarrow b}_{\bf p,k}}=-2\pi
\big|{\widetilde V}^{\rm b-ph}_{{\bf p};{\bf k},{\bf k}-{\bf p}}\big|^2
\\
&&\phantom{\frac{1}{\tau^{\rm ph-b}_{\bf p,k}}=}
\Big(1+ f^0_{\bf k-p} + f^0_{\bf k}\Big)
\delta\big(\varepsilon_{\bf k}+
\varepsilon_{\bf k-p}-\omega_{\bf p}\big).\nonumber
\end{eqnarray}
Similar consideration gives the phonon collision integral:
\begin{eqnarray}
\label{Stn_a}
&&St^{\rm ph}_{\bf p}[n,f]= \pi\sum_{\bf k}
\big|{\widetilde V}^{\rm b-ph}_{{\bf p};\frac{{\bf k}+{\bf p}}{2},
\frac{{\bf k}-{\bf p}}{2}}\big|^2\\
&&\phantom{St^{\rm ph}}
\big[(1+n_{\bf p}) f_{\frac{{\bf k}-{\bf p}}{2}}
f_{\frac{{\bf k}+{\bf p}}{2}}
- n_{\bf p}(1+ f_{\frac{{\bf k}-{\bf p}}{2}} )
(1+f_{\frac{{\bf k}+{\bf p}}{2}})\big]\nonumber\\
&&\phantom{St^{\rm ph}}
\times\delta\big(\varepsilon_{\frac{{\bf k}-{\bf p}}{2}}+
\varepsilon_{\frac{{\bf k}+{\bf p}}{2}}-\omega_{\bf p}\big),\nonumber
\end{eqnarray}
where the factor of $2$ has been removed to avoid double counting of the
final states and the symmetrized notations
for the momenta are used.
The drag rate of bosons on phonons  is given by
\begin{eqnarray}
\label{taubf_a}
&&St^{\rm ph}_{\bf k}[n^0,f^0+\delta f]\approx
-\sum_{\bf k}\frac{\delta f_{\bf k}}{\tau^{\rm b\rightarrow ph}_{\bf k,p}}
,\nonumber\\
&&\frac{1}{\tau^{\rm b\rightarrow ph}_{\bf k,p}}=-2\pi
\big|{\widetilde V}^{\rm b-ph}_{{\bf p};{\bf k},{\bf k}-{\bf p}}\big|^2 \big[f^0_{\bf k-p} - n^0_{\bf p}\big]\nonumber\\
&&\phantom{\frac{1}{\tau^{\rm b-ph}_{\bf k,p}}=-2\big|^2 }
\times
\delta\big(\varepsilon_{\bf k}+
\varepsilon_{\bf k-p}-\omega_{\bf p}\big).
\end{eqnarray}
After some algebra, the drag conductivity
from bosons on phonons and phonons on bosons turn out to be identical
and yield the total thermal conductivity of (\ref{kappaphf4}).

\subsection{Drag conductivity in a limiting case}

Here we derive the thermal
drag conductivity under two main assumptions: the boson
scattering times are independent of the momentum and those
of the phonons  are determined entirely by its
interaction with bosons.
We consider linear energy spectrum for both
bosons and phonons which are given by $\varepsilon_{\bf k} = v |{\bf k}|$
and $\omega_{\bf p} = u |{\bf p}|$, respectively. Thus the drag conductivity of Eq.~(\ref{kappatotal}) reduces to,
\begin{eqnarray}
\label{kappatotal-App}
&&\kappa_{\rm drag} = -\frac{4\pi u^2 v^2\tau^{\rm b}}{T}\sum_{\bf k,p}
{\bf k}^x\,{\bf p}^x\,\tau^{\rm ph}_{\bf p}\,
\big|V^{\rm b-ph}_{{\bf p};{\bf k},{\bf k}-{\bf p}}\big|^2\nonumber\\
&&\phantom{\kappa_{\rm drag} =}
\times\frac{\partial n^0_{\bf p}}{\partial \omega}
\big[f^0_{\bf k-p} -f^0_{\bf k}\big]\\
&&
\times
\bigg[
\delta\big(\varepsilon_{\bf k}-
\varepsilon_{\bf k-p}-\omega_{\bf p}\big)-
\delta\big(\varepsilon_{\bf k}-
\varepsilon_{\bf k-p}+\omega_{\bf p}\big)
\bigg].\nonumber
\end{eqnarray}
The phonon scattering time in (\ref{kappatotal-App}) can be obtained from the
phonon-boson collision integral in a standard way, similar to the derivation of the drag rates in Appendix~\ref{app_A}.1. Such a derivation yields,
\begin{eqnarray}
\label{tauphonon}
\frac{1}{\tau_{\bf p}^{\rm ph}}&=&2\pi \sum_{\bf k}
\big|V^{\rm b-ph}_{{\bf p};{\bf k},{\bf k}-{\bf p}}\big|^2
\times
\big[f^0_{\bf k-p} -f^0_{\bf k}\big]
\\
&&\bigg[
\delta\big(\varepsilon_{\bf k}-
\varepsilon_{\bf k-p}-\omega_{\bf p}\big)-
\delta\big(\varepsilon_{\bf k}-
\varepsilon_{\bf k-p}+\omega_{\bf p}\big)
\bigg].\nonumber
\end{eqnarray}
We now rewrite the drag term in a compact form,
\begin{eqnarray}
\label{kappacomp}
\kappa_{\rm drag} &=&   \frac{1}{3}v^2 \tau^{\rm b}\bar{C}_{\rm ph},
\end{eqnarray}
using the auxiliary function $\bar{C}_{\rm ph}$, which is a ``modified''
phonon specific heat given by,
\begin{eqnarray}
\label{Specific}
\bar{C}_{\rm ph}=- \frac{3}{T}\sum_{\bf p} \frac{\partial n^0_{\bf p}}{\partial \omega}
\left(u{\bf p}^x\right)^2\, F(p).
\end{eqnarray}
The   ${\bf k}$-integration is now hidden in another 
auxiliary function $F(p)$, which is given by
\begin{eqnarray}
\label{fp1}
&& {\bf p}^x F(p)=4\pi\tau^{\rm ph}_{\bf p}\sum_{\bf k}
{\bf k}^x\,
\big|V^{\rm b-ph}_{{\bf p};{\bf k},{\bf k}-{\bf p}}\big|^2
\times
\big[f^0_{\bf k-p} -f^0_{\bf k}\big]
\nonumber\\
&&\bigg[
\delta\big(\varepsilon_{\bf k}-
\varepsilon_{\bf k-p}-\omega_{\bf p}\big)-
\delta\big(\varepsilon_{\bf k}-
\varepsilon_{\bf k-p}+\omega_{\bf p}\big)
\bigg]\, .
\end{eqnarray}
Let us split the above expression into two terms
\begin{eqnarray}
\label{I1}
&&I_1=4\pi \tau_{\bf p}^{\rm ph} \sum_{\bf k}
{\bf k}^x
\big|V^{\rm b-ph}_{{\bf p};{\bf k},{\bf k}-{\bf p}}\big|^2
\big[f^0_{\bf k-p} -f^0_{\bf k}\big]
\nonumber\\
&&\times
\delta\big(\varepsilon_{\bf k}-
\varepsilon_{\bf k-p}-\omega_{\bf p}\big),
\end{eqnarray}
and
\begin{eqnarray}
\label{I2}
&&I_2=-4\pi  \tau_{\bf p}^{\rm ph} \sum_{\bf k}
{\bf k}^x \big|V^{\rm b-ph}_{{\bf p};{\bf k},{\bf k}-{\bf p}}\big|^2
\big[f^0_{\bf k-p} -f^0_{\bf k}\big]
\nonumber\\
&&\times
\delta\big(\varepsilon_{\bf k}-
\varepsilon_{\bf k-p}+\omega_{\bf p}\big).
\end{eqnarray}
In $I_2$  we make the change ${\bf k-p}\rightarrow{\bf -k} $,
\begin{eqnarray}
&&I_2= -4\pi  \tau_{\bf p}^{\rm ph} \sum_{\bf k}
({\bf k}^x -{\bf p}^x)\big|V^{\rm b-ph}_{{\bf p};{\bf p-k},{\bf -k}}\big|^2
\nonumber\\
&&\times
\big[f^0_{\bf k-p} -f^0_{\bf k}\big]
 \delta\big(\varepsilon_{\bf k}-
\varepsilon_{\bf k-p}-\omega_{\bf p}\big),
\end{eqnarray}
Thus, $\bar{C}_{\rm ph}$ in (\ref{Specific}) is given by,
\begin{eqnarray}
\label{fp2}
&& \bar{C}_{\rm ph}= - \frac{3 u^2}{T}\sum_{\bf p} \frac{\partial n^0_{\bf p}}{\partial \omega}
{\bf p}^x\,(I_1 +I_2)\nonumber\\
&&=- \frac{3 u^2}{T}\sum_{\bf p} \frac{\partial n^0_{\bf p}}{\partial \omega}
({\bf p}^x)^2\,\Big[4\pi\, \tau_{\bf p}^{\rm ph}\,  \sum_{\bf k}
\big|V^{\rm b-ph}_{{\bf p};{\bf k},{\bf k}-{\bf p}}\big|^2
\nonumber\\
&&\times
\big[f^0_{\bf k-p} -f^0_{\bf k}\big]
\delta\big(\varepsilon_{\bf k}-
\varepsilon_{\bf k-p}-\omega_{\bf p}\big)\Big],
\end{eqnarray}
where we have utilized the relations
$\big|V^{\rm b-ph}_{{\bf p};{\bf k},{\bf k}-{\bf p}}\big|^2 = \big|V^{\rm b-ph}_{{\bf -p};{\bf k-p},{\bf k}}\big|^2$
and $\tau_{\bf p}^{\rm ph}= \tau_{\bf -p}^{\rm ph}$. Performing similar
manipulations on $\tau_{\bf p}^{\rm ph}$ in (\ref{tauphonon}) 
one can see that the ${\bf k}$-integral in $\tau_{\bf p}^{\rm ph}$
cancels {\it exactly} the one in (\ref{fp2}). Thus,
\begin{eqnarray}
\label{fp3}
&& \bar{C}_{\rm ph}= - \frac{3 u^2}{T}\sum_{\bf p} \frac{\partial n^0_{\bf p}}{\partial \omega}
({\bf p}^x)^2 \equiv C_{\rm ph},
\end{eqnarray}
where $C_{\rm ph}$ is the phonon specific heat.
This is a rather remarkable result given the arbitrary form of boson-phonon interaction.
Thus, the thermal drag conductivity
expressed in terms of $\kappa_{b}=v^2\tau^{\rm b}C_{\rm b}/D_s$ is
\begin{eqnarray}
\frac{\kappa_{\rm drag}}{\kappa_{\rm b}} =\frac{D_s}{3}\cdot
 \frac{C_{\rm ph}}{C_{\rm b}},
\end{eqnarray}
where $C_{\rm b}$ is the boson specific heat, and $D_s$ is the dimensionality of the spin system.

We find that $\bar{C}_{\rm ph}\equiv C_{\rm ph}$ even for the scenario when 
3D-phonons interact with 1D bosons.
In this case, the 3D-phonon
momentum can be split into ${\bf p}={\bf p_\|}+{\bf p_\perp}$, the part
parallel to the 1D momentum of the boson ${\bf k}$ 
and the part perpendicular to it.  Thus, the boson occupation numbers in the 3D-1D case change to $(f^0_{\bf k-p_{\|}} -f^0_{\bf k})$
and the $\delta$-function changes to  $\delta\big(\varepsilon_{\bf k}-
\varepsilon_{\bf k-p_\|}+\omega_{\bf p}\big)$.  The momentum 
transformation to be used for this case is ${\bf k-p_\|}\rightarrow{\bf -k}$.

\section{Branch cut integration approach\label{sec:Analytic-continuation}\label{sec:BCA}}

Here we derive the results of Section \ref{Kubo}.A using a
different approach,
namely by converting the frequency summations in (\ref{IA}) and
(\ref{IB}) to the problem of integration along the branch cuts
of the Green's functions.  In
Section \ref{Kubo}.A we have analyzed in detail the diagram $A$ in
Fig.~\ref{diagrams}. Here we consider the derivation for the
diagram $B$. Carrying out the Matsubara
summation in Eq.~(\ref{IB}), obeying the location  of the branch
cuts of the Green's
functions and using the shorthand notation $\int_{x}=\int_{-\infty}^{+\infty}dx/2\pi$ leads to
\begin{eqnarray}
\lefteqn{\Pi_{B}(\mathbf{k},\mathbf{p},i\omega)=4\int_{x}\int_{y}\, n(x)n(y)\left[\vphantom{\sum}\right.}\nonumber \\
 & \vphantom{\sum^{{\textstyle A}}} & \left\{ -\mathrm{Im}[G_{\mathbf{k}}^{R}(y)\bar{G}_{\mathbf{p}}^{R}(x+y)]\right.\nonumber \\
 &  & \hphantom{aaa}\times G_{\mathbf{k}}(y-i\omega)\bar{G}_{\mathbf{p}}(x+y-i\omega)G''_{\mathbf{k}-\mathbf{p}}(-x)\nonumber \\
 &  & -\mathrm{Im}[G_{\mathbf{k}}^{R}(y)\bar{G}_{\mathbf{p}}^{R}(x+y)]\nonumber \\
 &  & \left.\hphantom{aaa}\times G_{\mathbf{k}}(y+i\omega)\bar{G}_{\mathbf{p}}(x+y+i\omega)G''_{\mathbf{k}-\mathbf{p}}(-x)\right\} \label{eq:6a}\\
 & \vphantom{\sum^{{\textstyle A}}} & +\left\{ \mathrm{Im}[G_{\mathbf{k}}^{R}(y)G_{\mathbf{k}-\mathbf{p}}^{R}(y-x)]\right.\nonumber \\
 &  & \hphantom{aaa}\times G_{\mathbf{k}}(y-i\omega)\bar{G}_{\mathbf{p}}(x-i\omega)\bar{G}''_{\mathbf{p}}(x)\nonumber \\
 &  & +G''_{\mathbf{k}}(y)G_{\mathbf{k}}(y+i\omega)\nonumber \\
 &  & \left.\hphantom{aaa}\times G_{\mathbf{k}-\mathbf{p}}(y-x+i\omega)\bar{G}_{\mathbf{p}}(x-i\omega)\bar{G}''_{\mathbf{p}}(x)\right\} \label{eq:6b}\\
 & \vphantom{\sum^{{\textstyle A}}} & +\left\{ G''_{\mathbf{k}}(y)G_{\mathbf{k}}(y-i\omega)\right.\nonumber \\
 &  & \hphantom{aaa}\times G_{\mathbf{k}-\mathbf{p}}(y-x-i\omega)\bar{G}_{\mathbf{p}}(x+i\omega)\bar{G}''_{\mathbf{p}}(x)\nonumber \\
 &  & +\mathrm{Im}[G_{\mathbf{k}}^{R}(y)G_{\mathbf{k}-\mathbf{p}}^{R}(y-x)]\nonumber \\
 &  & \left.\left.\hphantom{aaa}\times G_{\mathbf{k}}(y+i\omega)\bar{G}_{\mathbf{p}}(x+i\omega)\bar{G}''_{\mathbf{p}}(x)\right\} \right]\label{eq:6c}\\
 & \vphantom{\sum^{{\textstyle A}}} & =\Pi_{B}^{a}(\mathbf{k},\mathbf{p},i\omega)+\Pi_{B}^{b+c}(\mathbf{k},\mathbf{p},i\omega)\hphantom{aaaaaaaaaaaaaaaa}
\end{eqnarray}
where $\stackrel{{\scriptscriptstyle(-)}}{G}{}_{\mathbf{k}}^{R}(x)=
\stackrel{{\scriptscriptstyle (-)\,\,}}{G_{\mathbf{k}}}(i\omega_{n}\rightarrow
x+i0^{+})=\stackrel{{\scriptscriptstyle
(-)\,\,}}{G'_{\mathbf{k}}}(x)+i\stackrel{{\scriptscriptstyle
(-)\,\,}}{G''_{\mathbf{k}}}(x)$ refers to the retarded Greens functions and
their decomposition into real ($'$) and imaginary ($''$) parts and
$n(\ldots)$ is
the Bose distribution function. The subscripts `a', `b', and `c' refer the
contributions to $\Pi_{B}(\mathbf{k},\mathbf{p},i\omega)$ which stem from the
curly brackets labeled by Eqs. (\ref{eq:6a}), (\ref{eq:6b}), and
(\ref{eq:6c}). For the DC heat conductivity we need
\begin{equation}
\Pi_{B}(\mathbf{k},\mathbf{p})=\lim_{\omega\rightarrow0}\frac{1}{\omega}\mathrm{Im}\left[\Pi_{B}(\mathbf{k},\mathbf{p},i\omega\rightarrow\omega+i0^{+})\right]\,.\label{eq:7}
\end{equation}
We first take this limit focusing on $\Pi_{B}^{a}$.
The variables of integration can
be substituted such as to express this limit in terms of a
derivative of the distribution function
\begin{eqnarray}
\Pi_{B}^{a}(\mathbf{k},\mathbf{p}) & = & 4\int_{x}\int_{y}n(x)\frac{\partial n(y)}{\partial y}\nonumber \\
 & \vphantom{\sum_{{\displaystyle \prod}}} & \times\left\{ \mathrm{Im}[G_{\mathbf{k}}^{R}(y)\bar{G}_{\mathbf{p}}^{R}(x+y)]\right\} ^{2}G''_{\mathbf{k}-\mathbf{p}}(-x)\nonumber \\
 & \equiv & 4\int_{x}\int_{y}n(x)\frac{\partial n(y)}{\partial y}\, R_{a}(x,y,\mathbf{k},\mathbf{p})\,,\hphantom{aaaa}\label{eq:8}
\end{eqnarray}
where the abbreviation $R_{a}(x,y,\mathbf{k},\mathbf{p})$ has been defined. A
similar substitution cannot be achieved for the contributions from
Eqs. (\ref{eq:6b}), and (\ref{eq:6c}).
Instead, we expand the Greens functions to
lowest order in $\omega$
\begin{eqnarray}
\lefteqn{\Pi_{B}^{b+c}(\mathbf{k},\mathbf{p})=4\int_{x}\int_{y}\, n(x)n(y)\frac{{\displaystyle \partial}}{{\displaystyle \partial\omega}}\left[\vphantom{\sum}\right.}\nonumber \\
 &  & G''_{k}(y)\bar{G}''_{p}(x)\nonumber \\
 &  & \times\left\{ \mathrm{Im}[G_{k}^{R}(y+\omega)G_{k-p}^{R}(y-x+\omega)\bar{G}_{p}^{A}(x-\omega)]\right.\nonumber \\
 &  & \left.-\mathrm{Im}[G_{k}^{R}(y-\omega)G_{k-p}^{R}(y-x-\omega)\bar{G}_{p}^{A}(x+\omega)]\right\} \nonumber \\
 &  & +\bar{G}''_{p}(x)\mathrm{Im}[G_{k}^{R}(y)G_{k-p}^{R}(y-x)]\nonumber \\
 &  & \times\left\{ \mathrm{Im}[G_{k}^{R}(y+\omega)\bar{G}_{p}^{R}(x+\omega)]\right.\nonumber \\
 & \vphantom{\sum_{{\displaystyle \prod}}} & \left.\left.\left.-\mathrm{Im}[G_{k}^{R}(y-\omega)\bar{G}_{p}^{R}(x-\omega)]\right\} \right]\right|_{{\textstyle \omega=0}}\nonumber \\
 & \equiv & 4\int_{x}\int_{y}\, n(x)n(y)\, R_{b+c}(x,y,\mathbf{k},\mathbf{p})\,,\hphantom{aaaaaaaaaaaa}\label{eq:9}
\end{eqnarray}
As in Eqs. (\ref{A1}), and (\ref{A2}) we introduce a phenomenological, momentum
dependent one-particle self-energy for the bosons {[}phonons]
\begin{equation}
G[\bar{G}](z,\mathbf{k})=\frac{1}{z-\varepsilon_{\mathbf{k}}[\omega_{\mathbf{k}}]+ir_{\mathbf{k}}[s_{\mathbf{k}}]sgn(\mathrm{Im}(z))}\,,\label{eq:10}
\end{equation}
where $[]$-bracketed terms refer to phonons and $z$ is complex.  Inserting this
into Eq.~(\ref{eq:8}), $R_{a}(x,y,\mathbf{k},\mathbf{p})$ turns into a rational
function
\begin{eqnarray}
\lefteqn{R_{a}(x,y,\mathbf{k},\mathbf{p})=\frac{P_{a}(x,y,\mathbf{k},\mathbf{p})}{Q_{a}(x,y,\mathbf{k},\mathbf{p})}}\label{eq:11}\\
 & \vphantom{\sum^{{\textstyle A}}} & P_{a}(x,y,\mathbf{k},\mathbf{p})=-r_{\mathbf{k}-\mathbf{p}}[r_{\mathbf{k}}\omega_{\mathbf{p}}(x+y)+s_{\mathbf{p}}\varepsilon_{\mathbf{k}}(y)]^{2}\nonumber \\
 & \vphantom{\sum^{{\textstyle A}}} & Q_{a}(x,y,\mathbf{k},\mathbf{p})=[s_{\mathbf{p}}^{2}+\omega_{\mathbf{p}}(x+y)^{2}]^{2}\nonumber \\
 &  & \hphantom{aaaa}\times[r_{\mathbf{k}-\mathbf{p}}^{2}+\varepsilon_{\mathbf{k}-\mathbf{p}}(-x)^{2}][r_{\mathbf{k}}^{2}+\varepsilon_{\mathbf{k}}(y)^{2}]^{2}\,,\hphantom{aaaaaaa}\nonumber
\end{eqnarray}
where the abbreviations $\omega_{\mathbf{k}}(x)=x-\omega_{\mathbf{k}}$ and
$\varepsilon_{\mathbf{k}}(y)=y-\varepsilon_{\mathbf{k}}$ are used. Note that
$\omega_{\mathbf{k}}(x+y)=x+\omega_{\mathbf{k}}(y)$.  Similarly, in
Eq.~(\ref{eq:9}) $R_{b+c}(x,y,\mathbf{k},\mathbf{p})$ turns into
\begin{eqnarray}
\lefteqn{R_{b+c}(x,y,\mathbf{k},\mathbf{p})=\frac{P_{b+c}(x,y,\mathbf{k},\mathbf{p})}{Q_{b+c}(x,y,\mathbf{k},\mathbf{p})}}\label{eq:12}\\
 & \vphantom{\sum^{{\textstyle A}}} & P_{b+c}(x,y,\mathbf{k},\mathbf{p})=2s_{\mathbf{p}}\left\{ -2r_{\mathbf{k}}r_{\mathbf{k}-\mathbf{p}}[\omega_{\mathbf{p}}(x)^{2}+s_{\mathbf{p}}^{2}]\right.\nonumber \\
 &  & \times[r_{\mathbf{k}}^{2}+\varepsilon_{\mathbf{k}}(y)^{2}]\{r_{\mathbf{k}-\mathbf{p}}[\omega_{\mathbf{p}}(x)r_{\mathbf{k}}-s_{\mathbf{p}}\varepsilon_{\mathbf{k}}(y)]-[r_{\mathbf{k}}s_{\mathbf{p}}\nonumber \\
 &  & +\omega_{\mathbf{p}}(x)\varepsilon_{\mathbf{k}}(y)]\varepsilon_{\mathbf{k}-\mathbf{p}}(y-x)\}+[r_{\mathbf{k}-\mathbf{p}}^{2}+\varepsilon_{\mathbf{k}-\mathbf{p}}(y-x)^{2}]\nonumber \\
 &  & \times\left(-(r_{\mathbf{k}}-r_{\mathbf{k}-\mathbf{p}})s_{\mathbf{p}}[\omega_{\mathbf{p}}(x)^{2}+s_{\mathbf{p}}^{2}]\varepsilon_{\mathbf{k}}(y)[r_{\mathbf{k}}^{2}+\varepsilon_{\mathbf{k}}(y)^{2}]\right.\nonumber \\
 &  & +\omega_{\mathbf{p}}(x)s_{\mathbf{p}}\{-2r_{\mathbf{k}}^{4}r_{\mathbf{k}-\mathbf{p}}+r_{\mathbf{k}}^{4}s_{\mathbf{p}}-r_{\mathbf{k}}^{3}r_{\mathbf{k}-\mathbf{p}}s_{\mathbf{p}}\nonumber \\
 &  & +r_{\mathbf{k}}^{2}s_{\mathbf{p}}\varepsilon_{\mathbf{k}}(y)^{2}+3r_{\mathbf{k}}r_{\mathbf{k}-\mathbf{p}}s_{\mathbf{p}}\varepsilon_{\mathbf{k}}(y)^{2}+2r_{\mathbf{k}-\mathbf{p}}\varepsilon_{\mathbf{k}}(y)^{4}\nonumber \\
 &  & +4r_{\mathbf{k}}\varepsilon_{\mathbf{k}}(y)[r_{\mathbf{k}}(r_{\mathbf{k}}+s_{\mathbf{p}})+\varepsilon_{\mathbf{k}}(y)^{2}]\varepsilon_{\mathbf{k}-\mathbf{p}}(y-x)\}\nonumber \\
 &  & +\omega_{\mathbf{p}}(x)^{3}r_{\mathbf{k}}\{r_{\mathbf{k}}^{3}-r_{\mathbf{k}}^{2}r_{\mathbf{k}-\mathbf{p}}+3r_{\mathbf{k}-\mathbf{p}}\varepsilon_{\mathbf{k}}(y)^{2}\nonumber \\
 &  & \left.\left.+r_{\mathbf{k}}\varepsilon_{\mathbf{k}}(y)[\varepsilon_{\mathbf{k}}(y)+4\varepsilon_{\mathbf{k}-\mathbf{p}}(y-x)]\}\right)\right\} \nonumber \\
 & \vphantom{\sum^{{\textstyle A}}} & Q_{b+c}(x,y,\mathbf{k},\mathbf{p})=[s_{\mathbf{p}}^{2}+\omega_{\mathbf{p}}^{2}(x)]^{3}[r_{\mathbf{k}}^{2}+\varepsilon_{\mathbf{k}}^{2}(y)]^{3}\nonumber \\
 &  & \hphantom{Q_{b+c}(x,y,\mathbf{k},\mathbf{p})}\times[r_{\mathbf{k}-\mathbf{p}}^{2}+\varepsilon_{\mathbf{k}-\mathbf{p}}^{2}(y-x)]^{2}\nonumber
\end{eqnarray}
$\Pi_{B}^{a(b+c)}(\mathbf{k},\mathbf{p})$ will be evaluated assuming, as before,
that the bosons and phonons are quasiparticles with
$r_{\mathbf{k}}(s_{\mathbf{k}})\ll\varepsilon_{\mathbf{k}}(\omega_{\mathbf{k}})$.
In that case, expressions valid to leading order in
$r_{\mathbf{k}}(s_{\mathbf{k}})/\varepsilon_{\mathbf{k}}(\omega_{\mathbf{k}})$ for Eqs.~(\ref{eq:8}), and (\ref{eq:9}) are obtained from the residues of
$R_{a(b+c)}(x,y,\mathbf{k},\mathbf{p})$ \emph{alone}, while \emph{assuming} the
distribution functions to be holomorphic and retaining only their lowest-order
non-vanishing derivatives. Moreover, any imaginary part of the arguments of the
distribution functions arising in that process can be dropped. Since the poles
of $R_{a(b+c)}(x,y,\mathbf{k},\mathbf{p})$ stem from quadratic equations at
most, this calculation can be done analytically. We emphasize that the
proper evaluation of the higher-order contributions in
$r_{\mathbf{k}}(s_{\mathbf{k}})/\varepsilon_{\mathbf{k}}(\omega_{\mathbf{k}})$
to $\Pi_{B}^{a(b+c)}(\mathbf{k},\mathbf{p})$ would require a treatment of the
pole structure of the Bose distribution functions and their
derivatives. Analytically this is not feasible given
$R_{a(b+c)}(x,y,\mathbf{k},\mathbf{p})$.  This also implies that the
``non-Boltzmann'' terms of Appendix \ref{sec:NBC} are \emph{not} a systematic
account of \emph{all} next-leading order corrections. The leading-order analytic
calculation is tedious but straightforward. After some algebra we arrive at
\begin{eqnarray}
\lefteqn{\Pi_{B}^{a}(\mathbf{k},\mathbf{p})=-\frac{1}{2r_{\mathbf{k}}s_{\mathbf{p}}}\frac{\partial n(\varepsilon_{\mathbf{k}})}{\partial\varepsilon_{\mathbf{k}}}n(-\varepsilon_{\mathbf{k}-\mathbf{p}})\frac{\eta_{\mathbf{kp}}}{(\eta_{\mathbf{kp}}^{2}+e_{\mathbf{kp}}^{2})}}\nonumber \\
 &  & \Pi_{B}^{b+c}(\mathbf{k},\mathbf{p})=\frac{1}{2r_{\mathbf{k}}s_{\mathbf{p}}}\frac{\partial n(\varepsilon_{\mathbf{k}})}{\partial\varepsilon_{\mathbf{k}}}n(\omega_{\mathbf{p}})\frac{\eta_{\mathbf{kp}}}{(\eta_{\mathbf{kp}}^{2}+e_{\mathbf{kp}}^{2})}\,,\hphantom{aaaaa}\label{eq:13}
\end{eqnarray}
where $\eta_{\mathbf{kp}}=r_{\mathbf{k}}+r_{\mathbf{k}-\mathbf{p}}+s_{\mathbf{p}}$ and
$e_{\mathbf{kp}}=\omega_{\mathbf{p}}-\varepsilon_{\mathbf{k}}+\varepsilon_{\mathbf{k}-\mathbf{p}}$.
Thus, the rightmost fraction in both expressions
can be approximated by
$\pi\delta(\omega_{\mathbf{p}}-\varepsilon_{\mathbf{k}}+\varepsilon_{\mathbf{k}-\mathbf{p}})$.
The corresponding constraint
$\omega_{\mathbf{p}}\approx\varepsilon_{\mathbf{k}}-\varepsilon_{\mathbf{k}-\mathbf{p}}$
has also been used to rearrange the arguments of the distribution functions in
Eq.~(\ref{eq:13}). Diagram $A$ in Fig. \ref{diagrams} can be obtained directly
from the preceding derivation
by relabeling $\omega_{2}\rightarrow-\omega_{2}$ and by
realizing that
$-i\omega_{n}-\omega_{\mathbf{p}}+is_{\mathbf{p}}sgn(\mathrm{Im}(-i\omega_{n}))
=-[i\omega_{n}+\omega_{\mathbf{p}}+is_{\mathbf{p}}sgn(\mathrm{Im}(i\omega_{n}))]$.
I.e. $\chi_{A}$ can be obtained from Eq.~(\ref{eq:13}) simply by using the
symmetries: $s_{\mathbf{p}}=s_{-\mathbf{p}}$,
$\omega_{\mathbf{p}}\rightarrow\omega_{-\mathbf{p}}$,
$u_{\mathbf{p}}^{\mu}=-u_{-\mathbf{p}}^{\mu}$, and by replacing
$\omega_{\mathbf{p}}\rightarrow-\omega_{\mathbf{p}}$.  Since
$\kappa^{\mu\nu}=\kappa_{A}^{\mu\nu}+\kappa_{B}^{\mu\nu}$ and the total drag is
$\kappa_{\text{drag}}=\kappa_{12}+\kappa_{21}$, the final result is
\begin{eqnarray}
\lefteqn{\kappa_{\mathrm{drag}}=-\frac{4\pi}{T}
\sum_{\mathbf{k},\mathbf{p}}v_{\mathbf{k}}^{x}
\varepsilon_{\mathbf{k}}u_{\mathbf{p}}^{x}\omega_{\mathbf{p}}
|V_{\mathbf{p};\mathbf{k},\mathbf{k}-\mathbf{p}}^{\mathrm{b-ph}}|^{2}
\tau_{\mathbf{k}}^{b}\tau_{\mathbf{p}}^{ph}\times}\nonumber \\
 &  & \left\{ \frac{\partial f_{\mathbf{k}}^{0}}{\partial\varepsilon_{\mathbf{k}}}
[1+f_{\mathbf{k}-\mathbf{p}}^{0}+n_{\mathbf{p}}^{0}]
\delta(\varepsilon_{\mathbf{k}}-\varepsilon_{\mathbf{k}-\mathbf{p}}-\omega_{\mathbf{p}})
\right.\nonumber \\
 &  & \hphantom{aaa}\left.-
\frac{\partial f_{\mathbf{k}}^{0}}{\partial\varepsilon_{\mathbf{k}}}
[f_{\mathbf{k}-\mathbf{p}}^{0}-n_{\mathbf{p}}^{0}]
\delta(\varepsilon_{\mathbf{k}}-\varepsilon_{\mathbf{k}-\mathbf{p}}+\omega_{\mathbf{p}})
\right\} \,.\hphantom{aaaaa}\label{eq:16}
\end{eqnarray}
where we have renamed the Bose distribution functions with arguments
$\varepsilon_{{\bf k}}$ ($\omega_{\mathbf{k}}$) to $f_{\mathbf{{\bf k}}}^{0}$
($n_{\mathbf{k}}^{0}$), as in section \ref{Kubo}.A.
This result is \emph{identical} to Eq. (\ref{K_kappa_4}). Thus, both
technical approaches within the Kubo formalism yield the same answer.

\begin{widetext}
\section{Non-Boltzmann contributions\label{sec:NBC}}

In Section \ref{Kubo}.A we have discussed contributions from ${\rm
Im}\,\overline{\Pi}_{A}^{(4)}(\omega)$ and showed that they lead to the results  identical to the ones from Boltzmann theory. In the following we will discuss additional contributions to
drag thermal conductivity from the remaining terms of Eq. (\ref{IAB_4}), ${\rm
Im}\,\overline{\Pi}_{A}^{(m)}(\omega)$, $m=1,2$ and 3, which, however, are
subleading and can be safely neglected.  Evaluation of these terms is rather
cumbersome, and for illustrative purposes we focus only on ${\rm Im}\,
\overline{\Pi}_{A}^{(2)}(\omega)$, which is given by
\begin{eqnarray}
\label{B2_1}
&&{\rm Im}\, \overline{\Pi}_{A}^{(2)}({\bf k},{\bf p},\omega)=-\frac12
\int_{\varepsilon_1}\int_{\varepsilon_2}\int_{\varepsilon_3}
\int_{\varepsilon_5} A_{\bf k}(\varepsilon_1)
A_{\bf k}(\varepsilon_2)A_{\bf k-p}(\varepsilon_3)
{\bar A}_{\bf p}(\varepsilon_5+\omega)
{\bar A}_{\bf p}(\varepsilon_5)
\
P\frac{1}{\varepsilon_1-\varepsilon_2+\omega}\ \\
&& \ \ \
P\bigg[
\frac{(n_3-n_1)(n_{3-1}-n_5)}
{\omega+\varepsilon_5+\varepsilon_1-\varepsilon_3}-
\frac{(n_3-n_1)(n_{3-1}-n_{5+\omega})}
{\omega+\varepsilon_5+\varepsilon_1-\varepsilon_3}
+\frac{(n_3-n_2)(n_{3-2}-n_{5+\omega})}
{\varepsilon_5+\varepsilon_2-\varepsilon_3}-
\frac{(n_3-n_2)(n_{3-2}-n_5)}
{\varepsilon_5+\varepsilon_2-\varepsilon_3}
\bigg]
\,
,\nonumber
\end{eqnarray}
where $P$ stands for the principal value.
In the limit of zero-frequency, $\omega\rightarrow 0$, we obtain,
\begin{eqnarray}
\label{B2_2}
&&\frac{{\rm Im}\, \overline{\Pi}_{A}^{(2)}
({\bf k},{\bf p},\omega)}{\omega}\bigg|_{\omega=0}=
\int_{\varepsilon_1}\int_{\varepsilon_2}\int_{\varepsilon_3}
\int_{\varepsilon_5} A_{\bf k}(\varepsilon_1)
A_{\bf k}(\varepsilon_2)A_{\bf k-p}(\varepsilon_3)
\Big({\bar A}_{\bf p}(\varepsilon_5)\Big)^2
\
P\frac{1}{\varepsilon_1-\varepsilon_2}\
P\frac{(n_1-n_3)}{\varepsilon_1+\varepsilon_5-\varepsilon_3} \
\ \frac{\partial n_5}{\partial\varepsilon_5}
\,
.
\end{eqnarray}
Using the spectral representation (\ref{A1}) one can easily perform
integrations in $\varepsilon_2$. We further simplify the expression
by performing the $\varepsilon_1$ and $\varepsilon_3$
integrations on the terms containing $n_3$ and $n_1$, respectively, to obtain
\begin{eqnarray}
\label{B2_3}
&&\frac{{\rm Im}\, \overline{\Pi}_{A}^{(2)}
({\bf k},{\bf p},\omega)}{\omega}\bigg|_{\omega=0}=
\int_{\varepsilon_1}\int_{\varepsilon_5}  A_{\bf k}(\varepsilon_1)
\Big({\bar A}_{\bf p}(\varepsilon_5)\Big)^2
\bigg[ \frac{1}{2}
\frac{y_1}{y_1^2+r_{\bf k}^2}\
\frac{\partial n_1}{\partial\varepsilon_1}+
\Big(\frac{1}{2}\,\frac{\partial }{\partial y_1} \frac{y_1}{y_1^2+ r_{\bf k}^2} \Big)n_1
\bigg]
\frac{\partial n_5}{\partial\varepsilon_5}
\nonumber\\
&&\phantom{\frac{{\rm Im}\, I_{B}^{(\#2)}(\omega)}{\omega}\bigg|_{\omega=0}=} -
\int_{\varepsilon_3}
\int_{\varepsilon_5}
A_{\bf k-p}(\varepsilon_3)
\Big({\bar A}_{\bf p}(\varepsilon_5)\Big)^2 \,
\Big(\frac{1}{2}\,\frac{\partial }{\partial y} \frac{y}{y^2+ r_{\bf k}^2}\Big)
n_3 \frac{\partial n_5}{\partial\varepsilon_5}
\,
,\ \ \ \
\end{eqnarray}
where $y=(\varepsilon_{\bf k}+\varepsilon_5-\varepsilon_3)$
and $y_1=(\varepsilon_1+\varepsilon_5-\varepsilon_{\bf k-p})$. The contributions
from all three terms in the above expression are of the same order.
Consider contributions from the first term which is given by
\begin{eqnarray}
\label{B2_5}
\frac{1}{2}\int_{\varepsilon_1}\int_{\varepsilon_5}  A_{\bf k}(\varepsilon_1)
\Big({\bar A}_{\bf p}(\varepsilon_5)\Big)^2
\frac{y_1}{y_1^2+r_{\bf k}^2}\frac{\partial n_1}{\partial\varepsilon_1}\frac{\partial n_5}{\partial\varepsilon_5}\approx
\frac{\pi}{2}\,\frac{1}{r_{\bf k} s_{\bf p}}\, \frac{\partial n_{\bf p}^0}{\partial\omega}\,\frac{\partial f_{\bf k}^0}{\partial\varepsilon}\cdot
(\varepsilon_{\bf k}+\omega_{\bf p}-\varepsilon_{\bf k-p}) \cdot\delta(\varepsilon_{\bf k}+\omega_{\bf p}-\varepsilon_{\bf k-p}).
\end{eqnarray}
It appears that this expression contains the factor of the type
$x\cdot\delta(x)$, which implies $x\equiv 0$. However, under strict consideration, i.e., taking into account finite lifetime $r_{\bf k}$, $x$ is non-zero and is of the same order as the ``spread'' of the $\delta$-function ($x\sim r_{\bf k}$). From a
direct comparison of Eq.~(\ref{B2_5}) with Eq.~(\ref{B4_4}), we conclude that
contributions from (\ref{B2_5}) are smaller by the factor $(r_{\bf k} |\partial
f_{\bf k}^0/\partial\varepsilon|/f_{\bf k}^0)\sim r_{\bf k}/\varepsilon_{\bf k}\ll 1$.  Thus, the thermal
conductivity contributions from (\ref{B2_5}) and from the rest of the terms of
(\ref{B2_3}) can be neglected in comparison to the Boltzmann terms of
Eq.~(\ref{B4_4}). For the similar reason it is justified to use the delta-function form in Eq.~(\ref{K_kappa_4}) for the leading contributions, because the broadening in the spectral function only yields a subleading
correction of higher order in $r_{\bf k}[s_{\bf p}]$.

\end{widetext}

\end{document}